\documentclass[conference]{IEEEtran}
\IEEEoverridecommandlockouts
\usepackage{kotex}
\usepackage{url}
\usepackage{cite}
\usepackage{lipsum} 
\usepackage{graphicx}
\usepackage{amsmath, amsfonts, amssymb}
\usepackage{xcolor}
\usepackage{booktabs, multirow, array}
\ifCLASSOPTIONcompsoc
    \usepackage[caption=false, font=normalsize, labelfont=sf, textfont=sf]{subfig}
\else
\usepackage[caption=false, font=footnotesize]{subfig}

\usepackage{algorithm2e}
\RestyleAlgo{ruled}
\usepackage{etoolbox}
\makeatletter
\patchcmd{\@algocf@start}
  {-1.5em}
  {0pt}
  {}{}
\makeatother

\DeclareMathOperator*{\argmax}{arg\,max}

\begin{document}
\newcolumntype{B}{>{\centering\arraybackslash}p{0.05\textwidth}}

\title{Deep Polar Codes}


\author{   
Geon Choi, \IEEEmembership{Student Member, IEEE} and
 Namyoon Lee, \IEEEmembership{Senior Member, IEEE}
\thanks{
Geon Choi is with the Department of Electrical Engineering, POSTECH, Pohang 37673, South Korea
(e-mail: \mbox{simon03062@postech.ac.kr}).
}
\thanks{
Namyoon Lee is with the School of Electrical Engineering, Korea University, Seoul 02841, South Korea (e-mail: \mbox{namyoon@korea.ac.kr}).
}
}

\maketitle

\begin{abstract} 
In this paper, we introduce a novel class of pre-transformed polar codes, termed as \textit{deep polar codes.}  We first present a deep polar encoder that harnesses a series of multi-layered polar transformations with varying sizes. Our approach to encoding enables a low-complexity implementation while significantly enhancing the weight distribution of the code. Moreover, our encoding method offers flexibility in rate-profiling, embracing a wide range of code rates and blocklengths. Next, we put forth a low-complexity decoding algorithm called successive cancellation list with \textit{backpropagation parity checks} (SCL-BPC). This decoding algorithm leverages the parity check equations in the reverse process of the multi-layered pre-transformed encoding for SCL decoding. Additionally, we present a low-latency decoding algorithm that employs parallel-SCL decoding by treating partially pre-transformed bit patterns as additional frozen bits. Through simulations, we demonstrate that deep polar codes outperform existing pre-transformed polar codes in terms of block error rates across various code rates under short block lengths, while maintaining low encoding and decoding complexity. Furthermore, we show that concatenating deep polar codes with cyclic-redundancy-check codes can achieve the meta-converse bound of the finite block length capacity within 0.4 dB in some instances.




\end{abstract}


\section{Introduction}
 The quest for ultra-reliable low-latency communications (URLLC) in next-generation wireless systems remains relentless \cite{Wang-6G-Survey, tataria-6G-URLLC, zhang-6G-URLLC, david-6G-URLLC}. The concept of URLLC revolves around achieving lightning-fast data packet delivery within an incredibly short timeframe, such as 1 millisecond, while ensuring exceptional reliability, with packet error rates ranging from 10$^{-3}$ to 10$^{-6}$ \cite{Chen-URLLC}. The pursuit of URLLC remains unwavering for the next generation of wireless systems. The success of achieving ultra-low latency and high reliability is contingent on developing a cutting-edge channel coding technique capable of fast decoding and near-optimal performance in the finite-blocklength regime \cite{shirvanimoghaddam19, yue23}. While significant progress has been made in understanding finite-blocklength information theory \cite{polyanskiy-finite, Yang-finite-mimo-fading, Makki-finite-HARQ}, designing the optimal code in this context remains a formidable challenge due to the finite-blocklength constraint. In this paper, we make progress toward designing a new class of pre-transformed polar codes that can achieve the finite-blocklength capacity closely with a low-complexity encoder and decoder.
 

\subsection{Related Work}
 
Polar codes, invented by Arikan, represent a significant milestone as the first explicitly constructed error-correcting codes over binary input memoryless channels that achieve provably asymptotic capacity \cite{arikan-polar}. These codes utilize an explicit encoder and a successive cancellation (SC) decoder, boasting low complexity, approximately $\mathcal{O}(N \log N)$, where $N$ denotes the code blocklength \cite{arikan-polar}. However, it is essential to note that standard polar codes do not exhibit outstanding performance at short-to-moderate blocklengths compared to low-density parity-check (LDPC) or turbo codes \cite{Hui-5G-coding-comparison}. The reason is due to their inferior minimum distance and incomplete polarization. In order to address the challenges posed by incomplete polarization and enhance decoding robustness, SC-list (SCL) decoding was introduced in \cite{Tal-polar-SCL, Balatsoukas-SCL-log}. The SCL decoding has demonstrated superior performance compared to the standard SC decoder by increasing the list sizes of a decoder. Remarkably, as the list size grows, the performance of SCL decoding approaches that of maximum likelihood (ML) decoding. However, despite the utilization of ML decoding, the weight spectrum of polar codes is relatively inferior when compared to that of Reed-Muller (RM) codes \cite{mondelli-urbanke-polar2RM, li-tse-rm-polar, Abbe-Reed-Muller}.

Recently, there have been significant advancements in improving the weight spectrum of polar codes through pre-transform techniques at short-to-moderate blocklengths. One such approach involves concatenating polar codes with cyclic-redundancy-check (CRC) \cite{Tal-polar-SCL, Niu-CA-polar} and parity-check (PC) \cite{Wang-PCC-polar, Trifonov-polar-dynamic-frozen, Trifonov-polar-subcode, Zhang-PC-polar-Huawei}. In addition,  Arikan's PAC codes \cite{arikan-pac} leverage convolutional precoding as a pre-transformation. In the binary input additive white Gaussian noise (BI-AWGN) channel, it turns out that the PAC codes with RM rate-profiling can achieve dispersion bound very closely under Fano decoding or SCL decoding using large list sizes \cite{arikan-pac, moradi-pac, vardy-pac}. Notwithstanding the quasi-optimal performance, the major drawback of the PAC codes lies in their high decoding complexity. For instance, the tree-search-based sequential decoders (e.g., Fano or stack decoders) can result in considerable decoding delay at a low signal-to-noise ratio (SNR), which is not applicable for low latency scenarios. Also, the SCL decoding with a large list size increases hardware complexity considerably. To reduce the decoding complexity, some prior studies in \cite{Moradi-PAC-Fano-metric, Rowshan-PAC-Fano-metric} have put forth various extensions of the sequential decoding using improved metrics. Some notable prior studies in \cite{Liu-PAC-construction, Rowshan-PAC-construction, Moradi-pac-monte-carlo, Miloslavskaya-recursive-design, Chiu-PAC-consgruction} have proposed methods for constructing PAC codes with enhanced rate-profiling methods. Despite these efforts, finding the optimal rate-profiling method and the convolutional pre-transform at various rates and blocklengths remains an open problem.


A significant finding has recently emerged in pre-transformed polar codes, demonstrating that any upper-triangular pre-transformation applied to the polar codes does not reduce the minimum distance of the original polar codes \cite{li-pretransformed, li-pretransformed-average}. This finding offers an unified perspective on pre-transformed polar codes, encompassing various instances such as CRC-Aided (CA) polar (CA-polar), PC-polar, and PAC codes, all of which utilize upper-triangular transformation matrices for pre-transformation \cite{Niu-CA-polar, Wang-PCC-polar, Trifonov-polar-dynamic-frozen, Trifonov-polar-subcode, Zhang-PC-polar-Huawei, arikan-pac, Zhou-segmented-CApolar-SCL, Liang-segmented-BCH-CApolar-SCL, Gelincik-polar-row-merging}. This result sheds new light on the efficient construction of pre-transformed polar codes under an upper triangular structure of the precoding matrix. Despite remarkable recent advances in the pre-transformed polar codes, finding the optimal yet pragmatic pre-transformed polar codes remains exceptionally challenging. Such codes require (i) operating at rates close to finite-blocklength capacity, (ii) having low-complexity pre-transformation with an efficient rate-profiling, and (iii) having a low-complexity decoding algorithm.

    \subsection{Contributions}

     In this paper, we put forth a novel pre-transformed polar codes, referred to as \textit{deep polar codes.}  The main contributions of this work are summarized as follows:
     
    
        \begin{itemize}
    	
\item    	  We first present a novel successive encoding method for constructing deep polar codes. The deep polar encoder comprises a sequence of $L-1$ multi-layered pre-transformations, each with different sizes, followed by a regular polar transformation in the final layer. In each layer, a part of the information bits is encoded using the polar kernel matrix. The resulting transformed output is then fed into the partial input of the subsequent layer's encoder, along with additional information bits for the following layer encoding. Upon reaching the last layer, the encoder utilizes a regular polar transformation matrix to generate the final codewords. The proposed multi-layered pre-transformation method using the polar kernel matrices enhances the weight distribution of the resulting codes due to their upper triangular structures while enabling low-complexity encoding through a fast polar transformation algorithm.  
    	

%

        \item  We then propose a flexible rate-profiling method for deep polar encoding. By harnessing the polar transform matrix used in each layer encoding, the encoder independently constructs three index sets: i) information, ii) connection, and iii) frozen sets. The information set per layer consists of highly reliable (almost perfectly polarized to the capacity of one) bit-channel indices. Then, the connection set composes of less polarized bit-channel indices while maintaining a pre-defined minimum Hamming distance. The proposed rate-profiling method is flexible with respect to i) the number of layers, ii) code rates, and iii) blocklengths, because of its independent rate-profiling structure over the layers. 

    \item We present some construction examples of deep polar codes under symmetric binary erasure channels (BECs) with a blocklength of 32 and operating at two different rates. Through these examples, we provide an intuitive explanation of how the adoption of partial pre-transformation is sufficient to enhance the weight distribution of polar codes while enabling the design of a low-complexity decoder.

 \item  We introduce two computationally-efficient decoding algorithms for deep polar codes. The first algorithm is called SCL with a backpropagation parity check (SCL-BPC) decoder. The main idea of the SCL-BPC decoder is to leverage bit-wise parity check equations in the reverse process of successive deep polar encoding. This BPC mechanism plays a crucial role in pruning the list that fails BPC equations when performing SCL decoding. We also introduce a low-latency decoding algorithm called parallel-SCL decoder. The main idea of the parallel-SCL decoder is to perform multiple SCL decoders in parallel by treating different hypotheses of connection bit patterns as additional frozen bits.

\item We present simulation results evaluating the effectiveness of deep polar codes under BI-AWGN channels. When employing the SCL-BPC decoder with sufficient list sizes, deep polar codes are comparable with the PAC codes; both achieve the normal approximation bound tightly at rates of $R\in \{\frac{29}{128},\frac{64}{128}\}$. Remarkably, when using a list size of 8 for the SCL-type decoders, which is a more practically relevant scenario, the deep polar codes outperform both PAC and CA-polar codes across various code rates at blocklength of 128. This result advocates their superiority and potential for use in short packet transmissions. Our findings further indicate that deep polar codes exhibit superior decoding performance when employing the parallel-SCL decoder, making them a promising candidate for low-latency applications. Furthermore, we demonstrate that the CRC-aided deep polar (CA-deep polar) code with ML decoding achieves the meta-converse bound within 0.4 dB for blocklengths of 128 and 256 when the information size is 16. These results demonstrate that the CRC precoding can further boost the performance of the deep polar codes, leading to the finite-blocklength capacity closely.


    	
    \end{itemize}


%
%
%
%

\section{Preliminaries} 
In this section, we describe the preliminaries that are relevant to this work.

\subsection{Channel Coding System}

Consider an information vector ${\bf d}=[d_1,d_2,\ldots, d_K]\in \{0,1\}^K$, where $d_i$ represents independent and uniformly distributed bits over $\{0,1\}$ for $i\in [K]$. An encoder $\mathcal{E}:\{0,1\}^K\rightarrow \mathcal{X}^N$ maps the information vector ${\bf d}$ to a codeword ${\bf x} =[x_1,x_2\ldots, x_N]$ of length $N$, where $x_i$ is the $i$th element of the codeword, which is chosen from a binary alphabet set $\mathcal{X}$, i.e., $x_i\in \mathcal{X}$. The code rate $R$ is defined as the ratio of transmitted information bits to code blocklength, i.e., $R=\frac{K}{N}$.

Let $W: \mathcal{X}\rightarrow \mathcal{Y}$ be a binary input discrete memoryless channel (B-DMC) with output alphabet set $\mathcal{Y}$. The codeword ${\bf x}$ is transmitted over this B-DMC, resulting in an output sequence ${\bf y}=[y_1,y_2\ldots, y_N]$ of length $N$. A decoder $\mathcal{D}:\mathcal{Y}^N\rightarrow \{0,1\}^K$ produces an estimate of the message bits ${\bf \hat d}$. 

{\bf Symmetric channel capacity:}
 The symmetric capacity of B-DMC is defined as
\begin{align}
	I(W)=\sum_{x_i\in \{0,1\}}\sum_{y_i\in \mathcal{Y}}\frac{1}{2}W(y_i|x_i)\log_2\frac{W(y_i|x_i)}{\frac{1}{2}W(y_i|0)+\frac{1}{2}W(y_i|1)}. \nonumber
\end{align}
Further, the cutoff rate is computed as
\begin{align}
	R_0(W)=1-\log_2\left(1+ Z(W)\right),
\end{align}
where $Z(W)$ is the Bhattacharyya parameter defined as
\begin{align}
	Z(W)=\sum_{y_i\in \mathcal{Y}} \sqrt{W(y_i|0)W(y_i|1)}.
\end{align}

{\bf Block error probability:}
The block error rate (BLER) is defined as
\begin{align}
	 P(E) =\mathbb{P}[ {\bf d}\neq {\bf \hat d}].
\end{align}
For a linear block code with the minimum distance of ${\sf d}^{\rm min}$, the ML decoding performance over BI-AWGN channel is approximately given by \cite{costello-road2capacity}:
\begin{align}
    P_{\sf ML}(E) &\approx A_{ {\sf d}^{ \rm  min}} Q\left(\sqrt{ {\sf d}^{ \rm  min} {\sf SNR}}\right),
\end{align}
where $A_{ {\sf d}^{\rm min}} $ is the number of codewords with the minimum weight (or the nearest neighbors) and ${\sf  SNR}$ is the SNR of BI-AWGN channel, and
\begin{align}
    Q(x) = \frac{1}{\sqrt {2\pi}} \int_x^\infty e^{-y^2/2}{\rm d}y
\end{align}
is the standard Q-function that characterizes the tail probability of a Gaussian random variable. The two parameters, ${\sf d}^{ \rm  min}$ and $A_{{\sf d}^{\rm min}}$,  play a crucial role in determining the performance of a code under  ML decoding.



 \subsection{Polar Codes}

%


A polar code with parameters $(N,K,\mathcal{I})$ is characterized by a polar transform matrix of size $N=2^n$ and an index set $\mathcal{I}\subseteq [N]$. The polar transform matrix of size $N=2^n$ is obtained through the $n$th Kronecker power of a binary kernel matrix ${\bf G}_2=\begin{bmatrix}
1 & 0\\
1 & 1 
\end{bmatrix}$ as
 \begin{align}
	{\bf G}_N = {\bf G}_2^{\otimes n}.
\end{align}
The input vector of the encoder, ${\bf u}=[u_1,u_2,\ldots, u_N] \in \mathbb{F}_2^N$, is generated based on the given information set $\mathcal{I}$. In this process, the data vector carrying $K$ information bits, ${\bf d}$, is allocated to ${\bf u}_{\mathcal{I}}$. The remaining elements of ${\bf u}$, ${\bf u}_{\mathcal{I}^c}$, are assigned to zeros. Here, $\mathcal{I}^c=[N]/\mathcal{I}$ is referred to as the frozen bit set. This data assignment procedure is commonly known as rate-profiling. Finally, a polar codeword is constructed by multiplying ${\bf u}$ with ${\bf G}_N$ as 
\begin{align}
	{\bf x}^{\sf Polar}={\bf u}{\bf G}_N=\sum_{i\in \mathcal{I}} u_i {\bf g}_{N,i},
\end{align}
where ${\bf g}_{N,i}$ is the $i$th row vector of ${\bf G}_N$. According to the channel combining and splitting principle \cite{arikan-polar}, the $i$th bit-channel $W_N^{(i)}: \mathcal{X}\rightarrow \mathcal{Y}^N\times \mathcal{X}^{i-1}$, where $i\in [N]$, is defined as follows:
\begin{align}
    W_N^{(i)}\left({\bf y}, {\bf u}_{1:i-1} | u_i\right) = \sum_{{\bf u}_{i+1:N} \in \mathbb{F}_2^{N-i}} \frac{1}{2^{N-i}} W^N\left({\bf y}  | {\bf x}\right),
\end{align}
where $W^N\left({\bf y} | {\bf x}\right)$ is the  $N$ copies of B-DMCs and ${\bf u}_{a:b}=[u_a,u_{a+1},\ldots, u_{b}]$ for $a,b\in [N]$ and $a<b$. When $N$ is sufficiently large enough, the optimal rate-profiling is to select the indices having the capacity closed to one as
\begin{align}
	\mathcal{I}=\left\{ i\in [N] : I\left(W_N^{(i)} \right) =1-\epsilon \right\},
\end{align}
for small $\epsilon>0$.  This rate-profiling is sufficient to achieve the capacity under simple SC decoding \cite{arikan-polar}.  For a short blocklength regime, the information set consists of the indices that provide the $K$ most reliable bit-channels or the $K$ least bit-channel Bhattacharyya values.




 
\subsection{RM Codes}
Similar to polar codes, RM codes are constructed using ${\bf G}_N$, but with a different selection of rows. Instead of choosing rows based on the most reliable bit-channel capacity, RM codes simply select rows with the large weights. Let ${\sf wt}\left({\bf g}_{N,i}\right)$ represent the Hamming weight of the $i$th row of ${\bf G}_N$, where $i\in [N]$. For the $r$th-order RM code with length, $N=2^m$ and $0\leq r \leq m$, the information set is constructed by choosing the rows with a Hamming weight greater than or equal to $2^{m-r}$ as
\begin{align}
	\mathcal{I}_{{\sf RM}}=\{i\in [n]: {\sf wt}\left({\bf g}_{N,i} \right)\geq 2^{m-r}\}.
\end{align}
The encoder input vector ${\bf u}\in \mathbb{F}_2^N$ is formed by assigning the information vector ${\bf d}$ to ${\bf u}_{\mathcal{I}_{{\sf RM}}}$, while the remaining inputs $u_i=0$ for $i\notin \mathcal{I}_{{\sf RM}}$. Subsequently, a RM codeword is generated as 
\begin{align}
	{\bf x}^{\sf RM}={\bf u}{\bf G}_N=\sum_{i\in \mathcal{I}_{{\sf RM}}} u_i {\bf g}_{N,i}.
\end{align}
The RM code has a minimum distance of $2^{m-r}$ \cite{Abbe-Reed-Muller}. 


 
\subsection{Pre-transformed Polar Codes}

Polar and RM codes can be enhanced for error correction by utilizing a pre-transformation technique. A $(N,K,\mathcal{I},{\bf T})$ pre-transformed polar code consists of a binary pre-transformation matrix ${\bf T}\in \mathbb{F}_2^{N\times N}$ and an information set $\mathcal{I}$ for rate-profiling.  The construction of pre-transformed polar codes involves a two-stage encoding process. In the first stage, the information vector ${\bf d}\in \mathbb{F}^K$, carrying $K$ information bits, is inserted into the input vector ${\bf v}\in \mathbb{F}^N$ of the pre-transformation matrix. After assigning ${\bf v}_{\mathcal{I}}={\bf d}$ and ${\bf v}_{\mathcal{I}^c}={\bf 0}$, the first-stage encoding is performed by multiplying ${\bf v}$ with ${\bf T}$ as 
\begin{align}
	{\bf u}={\bf v}{\bf T}.
\end{align}
In the second stage encoding, the codeword ${\bf x}$ is generated by transforming the output of the first-stage encoding using ${\bf G}_N$ as 
\begin{align}
	{\bf x}={\bf u}{\bf G}_N={\bf v}{\bf T}{\bf G}_N.
\end{align}
The pre-transform matrix ${\bf T}$ and the rate-profiling index set $\mathcal{I}$ are required to be jointly optimized to enhance the decoding performance of finite-length polar codes. In a recent study \cite{li-pretransformed}, it was demonstrated that using an upper-triangular matrix ${\bf T}\in \mathbb{F}_2^{N\times N}$ with non-zero diagonal elements for the pre-transform guarantees to generate codewords with a minimum distance at least as large as that of polar codes (i.e., ${\bf T} = {\bf I}$). The PAC codes are representative examples of such pre-transformed polar codes, utilizing the upper-triangular Toeplitz matrix ${\bf T}$ as a pre-transformed matrix involving a convolution operation. However, determining the optimal information set $\mathcal{I}$ for the given ${\bf T}$ remains an unresolved challenge.

\section{deep polar Codes}
In this section, we present deep polar codes, a family of pre-transformed polar codes.  
 
\subsection{Encoding}
\begin{figure*}[t]
\centering
\includegraphics[width=1.7\columnwidth]{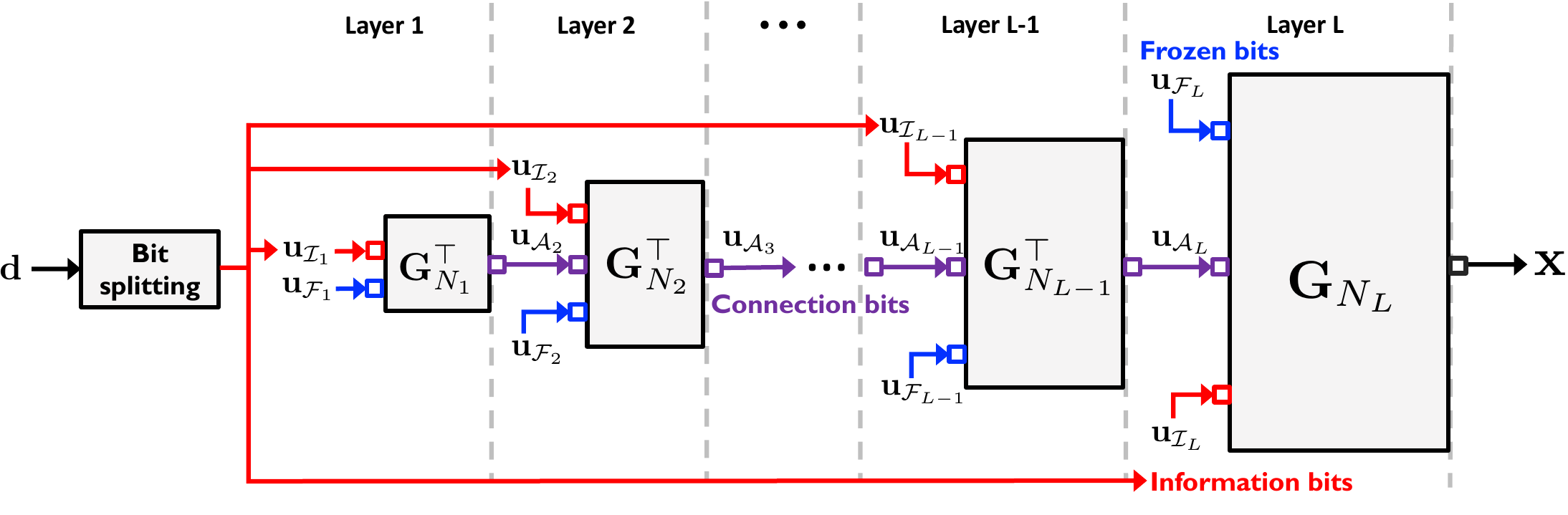}
\caption{An illustration of the proposed deep polar encoder with $L$ layers.}
\label{fig:encoder}
\end{figure*}

A  $\left(N,K, \{\mathcal{I}_{\ell}\}_{\ell=1}^L, \{\mathcal{A}_{\ell}\}_{\ell=1}^L, \{{\bf T}_{\ell}\}_{\ell=1}^L\}\right)$ deep polar code is defined with the following parameters:
\begin{itemize}
	\item  i) $L$ transformation matrices ${\bf T}_{\ell}\in \mathbb{F}_2^{N_{\ell}\times N_{\ell}}$,
	\item ii) $L$ information sets $\{\mathcal{I}_1,\mathcal{I}_2,\ldots, \mathcal{I}_{L}\}$, and
	\item iii) $L$ connection sets $\{\mathcal{A}_1,\mathcal{A}_2,\ldots, \mathcal{A}_{L}\}$.
\end{itemize}
 For given these parameters, the deep polar encoder performs  information bit splitting and successive encoding.
 
  \vspace{0.2cm}
 {\bf Information bit splitting and mapping:} 
The information vector ${\bf d} \in \mathbb{F}_2^K$ carrying $K$ information bits is splitted into $L$ information sub-vectors ${\bf d}_{\ell}$, each with size of $K_{\ell}=|\mathcal{I}_{\ell}| (<N_{\ell})$ for ${\ell}\in [L]$ and $\sum_{\ell=1}^L K_{\ell}=K$. 

 Let ${\bf u}_{\ell}=[u_{\ell,1},u_{\ell,2},\ldots, u_{\ell, N_{\ell}}]\in \mathbb{F}_2^{N_{\ell}}$ be the input vector of layer $\ell$ with length $N_{\ell}$ for $\ell\in [L]$. The index set of the $\ell$th layer is partitioned into three non-overlapped sub-index sets as
 \begin{align}
 	[N_{\ell}] =\mathcal{I}_{\ell}\cup\mathcal{A}_{\ell}\cup \mathcal{F}_{\ell},
 \end{align}
 where $\mathcal{F}_{\ell}=[N_{\ell}]/\{\mathcal{I}_{\ell}\cup \mathcal{A}_{\ell}\}$ is the frozen bit set of layer $\ell$ and $\mathcal{I}_{\ell}\cap \mathcal{A}_{\ell}=\phi$.  The information vector of the $\ell$th layer, ${\bf d}_{\ell}\in \mathbb{F}_2^{K_{\ell}}$, is assigned to ${\bf u}_{\mathcal{I}_{\ell}}$. Meanwhile, the frozen bits are assigned to ${\bf u}_{\mathcal{F}_{\ell}}={\bf 0}$, where $\mathcal{F}_{\ell}=[N_{\ell}]/\{\mathcal{I}_{\ell}\cup \mathcal{A}_{\ell}\}$. 
 
 \vspace{0.2cm}
{\bf Successive encoding:} As depicted in Fig.~\ref{fig:encoder}, an $L$-layered deep polar code is constructed by $L$-stage successive encoding procedures. Let ${\bf T}_{\ell}\in \mathbb{F}_2^{N_{\ell}\times N_{\ell}}$ be the $\ell$th layered pre-transformation matrix where $N_{1}<N_{2}<\cdots <N_{L-1}<N_L$. Unlike the PAC or other PAC-like codes, our deep polar codes adopt the pre-transform matrix by multiplying the transpose of the polar transform matrix with length $N_{\ell}=2^{n_{\ell}}$ for $n_{\ell}\in \mathbb{Z}^{+}$ as
\begin{align}
	{\bf T}_{\ell}= {\bf G}_{N_{\ell}}^{\top}.
\end{align}
The primary feature is that the proposed pre-transformation matrix ${\bf T}_{\ell}= {\bf G}_{N_{\ell}}^{\top}$ not only facilitates easy calculations using fast polar transform but also has an upper triangular matrix structure. This upper triangular structure allows for improvement in the weight distribution of the code. 


 In the first layer, the encoder generates the input vector of the first layer encoding, ${\bf u}_{1}=[  {\bf u}_{\mathcal{I}_{1}},~{\bf u}_{\mathcal{F}_{1}} ]$ where $\mathcal{A}_1=\phi$. The encoder output of the first layer is generated by ${\bf G}_{N_{1}}^{\top}$ as
\begin{align}
	{\bf v}_1 = {\bf u}_1{\bf G}_{N_{1}}^{\top}.
\end{align}
In the second layer,  the output vector of the first layer encoding is assigned to the input of the second layer to the connection index set $\mathcal{A}_2$, i.e., ${\bf u}_{\mathcal{A}_2}={\bf v}_1$. Since ${\bf u}_{\mathcal{I}_2}={\bf d}_2$ and ${\bf u}_{\mathcal{F}_2}={\bf 0}$, the input vector of the second layer encoding is 
\begin{align}
	{\bf u}_{2}=[  {\bf u}_{\mathcal{I}_{2}},~ {\bf u}_{\mathcal{A}_{2}},~{\bf u}_{\mathcal{F}_{2}} ].
\end{align} 
By multiplying ${\bf u}_{2}$ with ${\bf G}_{N_{2}}^{\top}$, the output vector of the second layer encoding is obtained as
\begin{align}
	{\bf v}_2 = {\bf u}_2 {\bf G}_{N_{2}}^{\top}.
\end{align}
Similar to the second layer encoding, the $\ell$th layer encoder for $2< \ell \leq L$ takes the input vector
\begin{align}
	{\bf u}_{\ell}=[  {\bf u}_{\mathcal{I}_{\ell}},~ {\bf u}_{\mathcal{A}_{\ell}},~{\bf u}_{\mathcal{F}_{\ell}} ],
\end{align} 
where ${\bf u}_{\mathcal{A}_{\ell}}={\bf u}_{\ell-1} {\bf G}_{N_{\ell-1}}^{\top}$, and generates the corresponding output vector by multiplying $ {\bf G}_{N_{\ell}}^{\top}$ or ${\bf G}_{N_L}$ as 
\begin{align}
{\bf v}_{\ell} &= 
\begin{cases}
	{\bf u}_{\ell}{\bf G}_{N_{\ell}}^{\top} & 2< \ell < L, \\
    {\bf u}_{L}{\bf G}_{N_{L}} & \ell = L. \\
\end{cases} \label{eq:encode}
\end{align}
For notational simplicity, we denote the output of the last layer encoder as the channel input ${\bf x}={\bf v}_L \in \mathbb{F}_2^{N_L}$.

\subsection{Rate-Profiling}
The selection of information and connection sets in all layers, $\mathcal{I}_{\ell}$ and $\mathcal{A}_{\ell}$, determines the error-correcting performance of deep polar codes. Unfortunately, finding the optimal set selection algorithm is very challenging. In this paper, we propose a flexible rate-profiling method that can support versatile code rates while guaranteeing an improved code weight distribution. The proposed rate-profiling method is to construct $\mathcal{I}_{\ell}$ and $\mathcal{A}_{\ell}$ for $\ell\in [L]$ individually across the layers.




\vspace{0.1cm}
{\bf Selection of $\mathcal{I}_{\ell}$ and $\mathcal{A}_{\ell}$:}  We explain how to select information and connection sets for the $\ell$th layer for $\ell\in \{1,\ldots, L\}$.  To construct $\mathcal{I}_{\ell}$ and $\mathcal{A}_{\ell}$ independently over the layers, suppose the channel input ${\bf x}_{\ell} \in \{0,1\}^{N_{\ell}}$ is formed by the transpose of polar transform ${\bf x}_{\ell}={\bf u}_{\ell}{\bf G}_{N_{\ell}}^{\top}$ for $\ell\in \{1,2,\ldots, L-1\}$ and the forward polar transform ${\bf x}_{L}={\bf u}_{L}{\bf G}_{N_{L}}$ for the last layer.  Then,  the codeword generated by layer $\ell$ encoder, ${\bf x}_{\ell}$, is assumed to be transmitted over an B-DMC, $W:  \mathcal{X}\rightarrow \mathcal{Y}$, which produces the channel output ${\bf y}_{\ell}$ for $\ell\in [L]$.  The $i$th bit-channel when using layer $\ell$ encoder is defined as
\begin{align}
    W_{N_{\ell}}^{(i)}\left({\bf y}_{\ell}, {\bf u}_{\ell,1:i-1} | u_{\ell,i}\right) = \sum_{{\bf u}_{\ell, i+1:N} \in \mathbb{F}_2^{N_{\ell}-i}} \frac{1}{2^{N_{\ell}-i}} W^{N_{\ell}}\left({\bf y}_{\ell}  | {\bf x}_{
    \ell}\right),
\end{align}
where $i\in [N_{\ell}]$, ${\bf u}_{\ell, a:b}=[u_{\ell,a},u_{\ell, a+1},\ldots, u_{\ell,b}]$ for $a,b\in [N_{\ell}]$ and $a<b$.  Let $I\left(W_{N_{\ell}}^{(i)}\right)$ be the $i$th bit-channel capacity.  

To construct $\mathcal{I}_{\ell}$ and $\mathcal{A}_{\ell}$, we first define an index set $\mathcal{R}_{\ell}$ according to the RM rate-profiling by selecting $i\in [N_{\ell}]$ such that the weight of the rows ${\bf G}_{N_{\ell}}$ is equal to or larger than the pre-determined ${\sf d}_{\ell}^{\rm min}$, i.e., 
 \begin{align}
	\mathcal{R}_{\ell}=\{i\in [N_{\ell}]:  {\sf wt}\left({\bf g}_{N_{\ell},i} \right)\geq {\sf d}^{\rm min}_{\ell} \},
\end{align}
where ${\sf d}^{\rm min}_{\ell}$ is a target minimum distance for the $\ell$th layer encoding. Next, we define an ordered index set of $\mathcal{R}_{\ell}$ as
\begin{align}
    \mathcal{\bar R}_{\ell}=\left\{i_1,i_2,\ldots, i_{|\mathcal{R}_{\ell}|}\right\},
\end{align}
 where $i_1$ is the index of the most reliable synthetic channel, i.e., $ I\left(W_{N_{\ell}}^{(i_1)}\right) \geq I\left(W_{N_{\ell}}^{(i_2)}\right)\geq \cdots \geq I\left(W_{N_{\ell}}^{(i_{|\mathcal{R}_{\ell}|})}\right)$. This ordered set can be constructed using the ordered Bhattacharyya values, i.e., $Z\left(W_{N_{\ell}}^{(i_1)}\right)\leq Z\left(W_{N_{\ell}}^{(i_2)}\right),\ldots \leq Z\left(W_{N_{\ell}}^{(i_{|\mathcal{B}_{\ell}|})}\right)$ or the values obtained from the density evolution with Gaussian approximation technique in \cite{Tal-polar-construction, Trifnov-polar-construction}.  Using the ordered index set, the information set $\mathcal{I}_{\ell}$ is generated by selecting the bit-channel indices that are approximately polarized to the capacity of one for given code length $N_{\ell}$:
\begin{align}
	\mathcal{I}_{\ell}=\left\{i\in \mathcal{\bar R}_{\ell}: I\left(W_{N_{\ell}}^{(i)}\right) \geq 1-\delta_{\ell} \right\},
\end{align}
where $\delta_{\ell}>0$ is chosen arbitrary small and $|\mathcal{I}_{\ell}|=K_{\ell}$.


 Next, we construct the connection set $\mathcal{A}_{\ell}$ as a subset of $\mathcal{\bar R}_{\ell}/\mathcal{I}_{\ell}$. Let $j_1,j_2,\ldots, j_{N_{\ell-1}} \in \mathcal{R}_{\ell}/\mathcal{I}_{\ell}$ be the indices that provide the highest $N_{\ell-1}$ bit-channel capacities in $\mathcal{R}_{\ell}/\mathcal{I}_{\ell}$ such that  $ I\left(W_{N_{\ell}}^{(j_1)}\right) \geq I\left(W_{N_{\ell}}^{(j_2)}\right)\geq \cdots \geq I\left(W_{N_{\ell}}^{(j_{N_{\ell-1}})}\right)$. Then,
\begin{align}
	\mathcal{ A}_{\ell}=\left\{j_1,j_2,\ldots, j_{N_{\ell-1}}\right\}.
\end{align}
 The frozen set of layer $\ell$ is defined as the collection of indices that are excluded in both the information and connection sets as
\begin{align}
	\mathcal{F}_{\ell}=[N_{\ell}]/\{\mathcal{I}_{\ell}\cup \mathcal{A}_{\ell}\}.
\end{align}
This rate-profiling method is performed independently over the layers. 


 \subsection{Remarks}
We provide some remarks on the encoding complexity, the superposition property, the minimum distance of the code, the concatenation with CRC codes, and the rate-compatibility.

 {\bf Encoding complexity:}  The proposed successive encoder composed of $L$ layers requires to take $L$ polar transformations, each with $N_{\ell}$ size. Since the computation of the polar transform with size $N_{\ell}$ needs the complexity of $\mathcal{O}(N_{\ell}\log_2 N_{\ell})$, the total encoding complexity boils down to
\begin{align}
	\mathcal{O}\left(\sum_{\ell=1}^LN_{\ell}\log_2 N_{\ell}\right).
\end{align}
It is worth mentioning that the encoding complexity can be comparable to that of the standard polar codes when $N_L$ is sufficiently larger than $N_{\ell}$ for $\ell\in [L-1]$. Choosing a small size of $N_{\ell}$ for $\ell\in [L-1]$ is a practical and effective strategy for reducing the encoding complexity.

\vspace{0.1cm}

{\bf Superposition codes :}  The deep polar code can be viewed with a lens through a superposition code of both the polar and the transformed polar codes as 
\begin{align}
 	{\bf x} = \underbrace{\sum_{j\in \mathcal{I}_{L}}u_{L,j} {\bf g}_{N_L, j} }_{{\bf x}_{\sf P}}+ \underbrace{\sum_{i\in \mathcal{A}_{L}}u_{L,i} {\bf g}_{N_L, i}}_{{\bf x}_{\sf TP}},
 \end{align}
where ${\bf x}_{\sf P}$ and  ${\bf x}_{\sf TP}$ represent polar and pre-transformed polar subcodewords, respectively. This superposition code interpretation provides a useful guideline for designing a low-complexity decoder while improving the weight distribution. Specifically, the pre-transformed subcodewords play a crucial role in improving the weight distribution of the deep polar code. The polar subcodewords facilitate to use a simple SC decoding because the information bits ${\bf u}_{\mathcal{I}_{L}}$ are sent through almost perfectly polarized bit-channels when the connection bits ${\bf u}_{\mathcal{A}_{L}}$ are treating as additional frozen bits. As a result, the deep polar codes offer a desirable balance between code performance and decoding complexity by strategically allocating information bit sizes between ${\bf u}_{\mathcal{I}_{L}}$ and ${\bf u}_{\mathcal{A}_{L}}$. For instance, when the blocklength goes to infinity, the encoder allocates all information bits to ${\bf u}_{\mathcal{I}_{L}}$ while $\mathcal{A}_{L}=\phi$; the deep polar codes boil down to a standard polar code.

 {\vspace{0.1cm}}

 {\bf The minimum distance and weight distribution:} The minimum distance of the deep polar codes is larger or equal to the minimum distance of layer $L$, ${\sf d}_{L}^{\rm min}$.  By the construction, the encoder chooses the row vectors of ${\bf G}_{N_L}$ with the weight larger than ${\sf d}_{L}^{\rm min}$, namely, 
 \begin{align}
{\sf wt}\left( {\bf g}_{N_L, j}\right)	\geq {\sf d}_{L}^{\rm min}
 \end{align}
for $j\in \mathcal{I}_L \cup \mathcal{A}_L$. Since the pre-transform matrix of layer $L-1$ holds the upper triangular structure, the minimum distance of the pre-transformed subcodewords ${\bf x}_{\sf TP}=\sum_{i\in \mathcal{A}_{L}}u_{L,i} {\bf g}_{N_L, i}$ cannot be less than ${\sf d}_{L}^{\rm min}$.

 

%

 {\vspace{0.1cm}}
 
 {\bf CRC-aided (CA) deep polar codes:} 
One possible extension of deep polar codes is to concatenate them with CRC codes. In this approach, the $K$-bit information vector ${\bf d}$ is precoded by using CRC generator polynomials, adding CRC bits. The length of the resulting information sequence becomes $K+K_{\rm CRC}$, where $K_{\rm CRC}$ represents the number of CRC bits. Alternatively, multiple CRC bits, each with short length, can be appended to the information bits of each layer, i.e., ${\bf u}_{\mathcal{I}_{\ell}}$ for $\ell\in [L]$ to improve the decoding performance. The effect of CRC code concatenation with the deep polar codes will be discussed in the simulation section.

 {\vspace{0.1cm}}
 
 {\bf Rate-compatibility:} 
A design of rate-compatible codes suitable for hybrid automatic repeat request (HARQ) is an important feature for practical communication systems. Our deep polar code possesses a rate-compatible property, thanks to its multi-layered successive encoding structure. Suppose a transmitter sends a codeword ${\bf x}$ and a receiver fails to decode the message bits ${\bf d}$. Then, in the next round of the transmission, the transmitter sends only the output of $L-1$ layer, ${\bf u}_{\mathcal{A}_L}$, with blocklength of $N_{L-1}$ for enabling HARQ communications. The receiver attempts to decode the message by using both the noisy connection bits received in the second round and the received signal in the first round. This re-transmission and decoding protocol can be iteratively applied to the first layer, enhancing the decoding performance while reducing the code rates from $R=\frac{K}{N_L}$ to $R=\frac{K}{\sum_{\ell=1}^LN_{\ell}}$. Notwithstanding the flexible rate-compatibility, a delicate code optimization is required to 
attain a high HAQR performance by carefully choosing the blocklengths $N_{\ell}$ and information bits $K_{\ell}$ for each layer $\ell\in \{1,2,\ldots,L-1 \}$. Solving this optimization problem remains as a promising future topic.

%
%
%


\section{Examples }
In this section, we provide some examples of the design of deep polar codes in a short blocklength regime to better understand the proposed encoding method.  Throughout the examples, we shall focus on a short packet transmission scenario with a length of $32$ over a BEC with an erasure probability of 0.5, i.e., $I(W)=0.5$ at two different rates $R\in \{\frac{11}{32},\frac{15}{32}\}$.

\subsection{Channel Polarization and Row Weights of ${\bf G}_{32}$}

\begin{figure}[t]
\centering
\includegraphics[width=1\columnwidth]{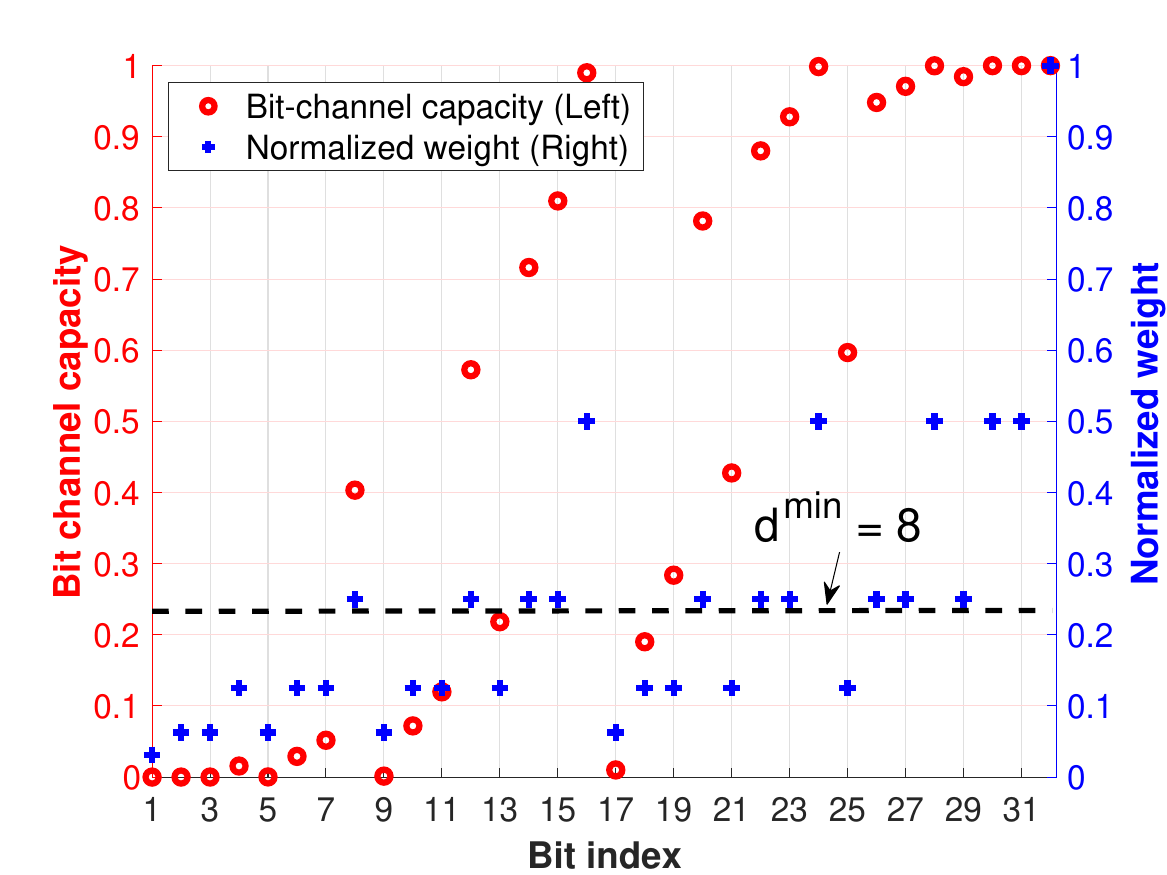}
\caption{Polarized bit-channel capacity (left) and normalized weight (right) for the BEC with $I(W)=0.5$.}
\label{fig:rateprofiling}
\end{figure}

For a deep polar code construction, it is important to understand the polarized bit-channel capacities and the row weight of ${\bf G}_{32}$. As shown in Fig. \ref{fig:rateprofiling}, the red circles indicate the bit-channel capacities for the BEC with $I(W)=0.5$, i.e., $I\left(W_{32}^{(i)}\right)$ for $i\in [32]$. The blue cross indicates the normalized weight of ${\bf G}_{32}$ according to index $i\in [32]$, i.e., $\frac{{\sf wt}({\bf g}_{32,i})}{32}$. The ordered index sets are mismatched according to bit-channel capacities and the normalized weights. For instance, $I\left(W_{32}^{(25)}\right)> I\left(W_{32}^{(12)}\right)$ but $\frac{{\sf wt}({\bf g}_{32,25})}{32} < \frac{{\sf wt}({\bf g}_{32,12})}{32}$.  When including the bit-index 25 in the information set to facilitate SC decoding, the minimum distance of the codewords is necessarily reduced. It becomes crucial to carefully select the bit-indices to improve the coding performance, considering both their weights and the bit-channel capacities. 


\subsection{Example 1}

This example explains how to construct a two-layered deep polar code with a code rate of $R=\frac{11}{32}$. The encoding process involves utilizing two transformation matrices: ${\bf G}_{8}^{\top}$ for layer 1 and ${\bf G}_{32}$ for layer 2.  Assuming that a target minimum distance of the second layer is ${\sf d}^{\rm min}_{2}=8$, we select the indices of rows in the matrix ${\bf G}_{32}$ whose weights are greater than or equal to ${\sf d}^{\rm min}_{2}=8$, i.e.,
 \begin{align}
	\mathcal{R}_{2}=\{8,12,14,15,16,20,22,23,24,26,27,28,29,30,31,32\}.\nonumber
\end{align}
Then, from Fig. \ref{fig:rateprofiling}, the information set for the second layer, a subset of $\mathcal{R}_{2}$, is chosen as
\begin{align}
	\mathcal{I}_2&=\left\{i\in \mathcal{R}_2: I\left(W_{32}^{(i)}\right) \geq 0.98 \right\}\nonumber\\
	& =\{32,31,30,28,24,16,29\},\nonumber
\end{align}
where $|\mathcal{I}_2|=K_2=7$.  Considering that $N_1=8$, we identify the eight bit-channel indices that yield the highest capacity in $\mathcal{R}_2/\mathcal{I}_2$ to form the connection set for the second layer as
  \begin{align}
	\mathcal{A}_2 =\{ 27,26,23,22,15,20,14,12\}.\nonumber
\end{align}
It is worth mentioning that the bit-channel index of $25$ was excluded in the connection set although its capacity is larger than that of the index of $12$, i.e.,  $I\left(W_{32}^{(25)}\right) \geq I\left(W_{32}^{(12)}\right)$. This is because ${\sf wt}\left({\bf g}_{32,25} \right)=4<{\sf wt}\left({\bf g}_{32,12} \right)=8$. The frozen set of the second layer becomes
\begin{align}
	\mathcal{F}_2=\{1,2,\ldots, 32\}/\{\mathcal{I}_2\cup\mathcal{A}_2\}.\nonumber
\end{align}
In the first layer, we identify the four indices that result in the highest bit-channel capacity using the polar transform matrix ${\bf G}_8^{\top}$, while ensuring that the minimum distance ${\sf d}^{\rm min}_1$ is greater than or equal to 4. The corresponding information and frozen sets are chosen as $\mathcal{I}_1=\{1,2,3,5\}$ and $\mathcal{F}_1=\{8,7,6,4\}$, respectively. The output vector of the first layer, denoted as ${\bf v}_1={\bf u}_{1}{\bf G}_8^{\top}$, is then connected to the input of connection set ${\bf u}_{\mathcal{A}_2}={\bf v}_1$.

\begin{table}
\centering
\caption{Comparison of the weight distributions }\label{table:WD}
\begin{tabular}{c|ccc|cccccc}
\toprule
{} & {} & $[32,11]$ & {} & {} & $[32,15]$  & {} \\ 
\cmidrule(lr){1-1}\cmidrule(lr){2-4}\cmidrule(lr){5-7}
{Weight} & Polar & RM-type & Proposed & Polar & RM-type & Proposed \\ 
\midrule
$0$  &  $1$        &  $1$     &  $1$          & $1$           &  $1$            &  $1$\\
$4$  &  $0$        &  $0$     &  $0$          & $8$           &  $0$            &  $0$\\
$8$  & $76$       &  $40$   &  $20$        & $444$       &  $364$        &  $300$\\
$12$ & $192$     &  $336$  &  $416$      & $6328$     &  $6720$      &  $6976$\\
$16$ &  $1510$  &  $1294$ &  $1174$    & $19206$   &  $18598$    &  $18214$\\ 
$20$ & $192$     &  $336$  &  $416$       & $6328$    &  $6720$      &  $6976$\\  
$24$ & $76$       &  $40$    &  $20$        & $444$       &  $364$       &  $300$\\
$28$ & $0$         &  $0$     &  $0$           & $8$          &  $0$           &  $0$\\
$32$ & $1$         &  $1$     &  $1$           & $1$          &  $1$           &  $1$\\
\bottomrule
\end{tabular}
\end{table}

\subsection{Example 2}
We present an additional example to construct an $\left(32,15, \{\mathcal{I}_1, \mathcal{I}_2\}, \{\phi, \mathcal{A}_2\}, \{{\bf G}_{4}^{\top}, {\bf G}_{32}\}\right)$ deep polar code. Applying the same approach as in Example 1, we first select the indices corresponding to the row vectors of ${\bf G}_{32}$ with a weight greater than 8 in an identical manner as  \begin{align}
 	\mathcal{R}_{2}=\{8,12,14,15,16,20,22,23,24,26,27,28,29,30,31,32\}.\nonumber
 \end{align}
We adopt the deep polar rate-profiling method by selecting the information and connection sets as
\begin{align}
	\mathcal{I}_2=\{15,16,22,23,24,26,27,28,29,30,31,32\}\nonumber
\end{align}
with $K_2=|\mathcal{I}_2|=12$
and
\begin{align}
	\mathcal{A}_2=\{8,12,14,20\}.\nonumber
\end{align}
Then, we choose the information set for the first layer as
\begin{align}
	\mathcal{I}_1=\{1,2,3\}\nonumber
\end{align}
with $K_1=|\mathcal{I}_{1}|  =3$.


\subsection{Comparisons with RM-type and Polar Codes}

We compare our deep polar codes in Examples 1 and 2 with existing RM and polar codes. We generate the polar codes for fair comparison by selecting the top $K\in \{11,15\}$ bit-channel indices that provide the highest capacities. For the RM code construction with information bit size $K\in \{11,15\}$,  we consider a subcodeword set of a $[32,16]$ RM code. Since the information set of the $[32,16]$ RM code is 
\begin{align}
	\mathcal{I}_{\sf RM}^{16} =\{i: {\sf wt}({\bf g}_{32,i})\geq 8\},
\end{align}
we choose the information set for $K=15$ as a subset of $ \mathcal{I}_{\sf RM}^{16}$, i.e., $\mathcal{I}_{\sf RM}^{15}\subseteq \mathcal{I}_{\sf RM}^{16}$. In particular, to optimize the code performance, we evaluate the weight distributions of 16 possible sub-codebooks of the $[32,16]$-RM code and select the best one with the smallest number of the codewords with the minimum weights.

\vspace{0.1cm}
{\bf Weight distribution:}
As shown in Table \ref{table:WD}, the proposed deep polar codes provide superior weight distributions than those of the RM and polar codes in both code rates. Specifically, deep polar codes have fewer codewords having the minimum weight than the RM codes while keeping the identical minimum distance of $8$ in both code rates. In addition, the deep polar code with a rate of $R=\frac{15}{32}$ can improve both the minimum distance and the number of codewords having the smallest weights.

\begin{figure}[t]
\centering
\includegraphics[width=0.9\columnwidth]{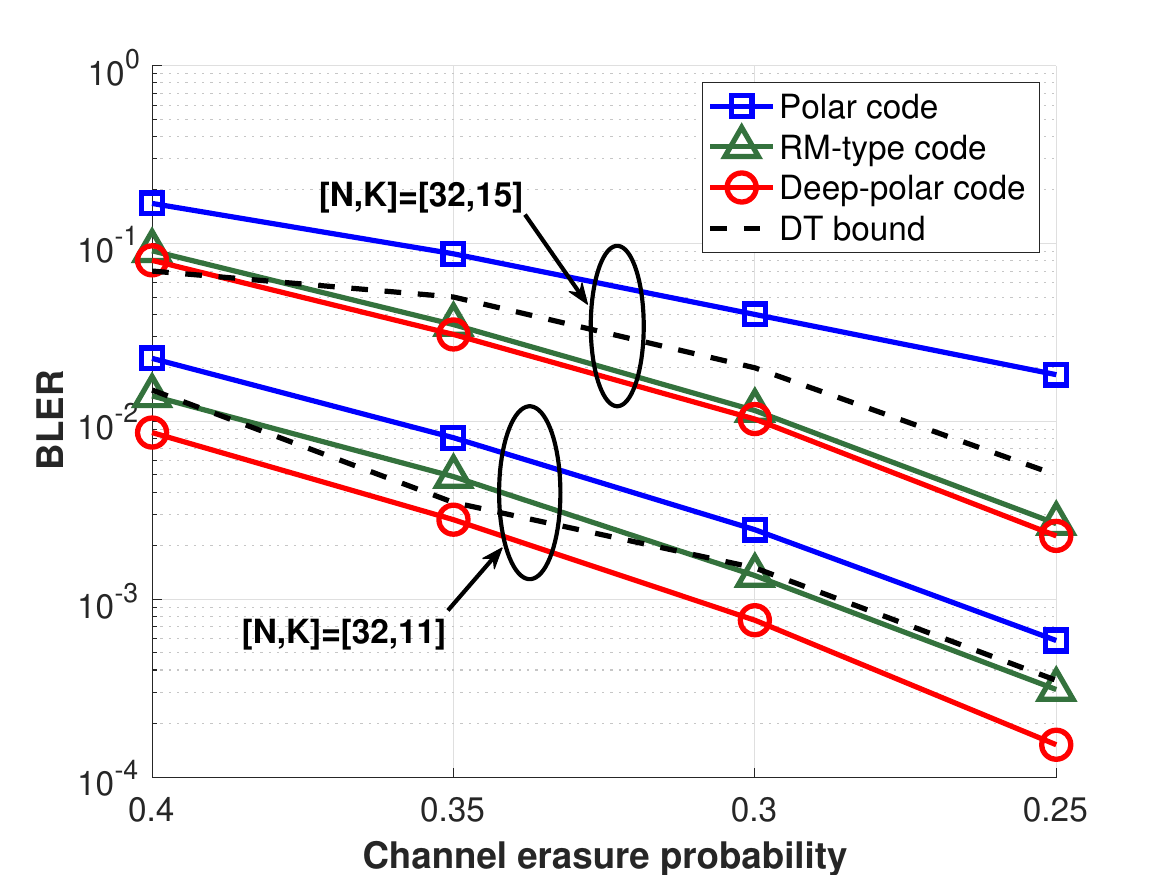}
\caption{BLER performance comparison for the BEC channel with $N=32$ and $K\in \{11,15\}$.}
\label{fig:BLER_BEC} \vspace{-0.5cm}
\end{figure}

{\bf BLER performance:}
To demonstrate the effect of the weight distribution improvement in the code design, we plot BLERs for the three codes under ML decoding as increasing the erasure probabilities of the BEC. As illustrated in Fig. \ref{fig:BLER_BEC}, the proposed deep polar provides noticeable BLER improvement compared to the RM and polar codes when $K=11$. This BLER gain stems from the considerable reduction of the codewords with the minimum weight. When $K=15$, the minimum distance of deep polar and RM codes is larger than the polar code, which leads to improved BELR performance. Although the deep polar and RM codes have identical minimum distances, the deep polar code slightly outperforms the RM code by reducing the codewords with the minimum weight, as shown in Table \ref{table:WD}. One remarkable result is that the deep polar codes can achieve better BLER performances than the dependence-testing (DT) bound, one of the strongest achievability bounds for the BEC in the finite blocklength regime. 


\vspace{-0.1cm}

\vspace{0.1cm}

{\bf Pre-transform using a sub-upper triangular matrix:} In contrast to conventional pre-transformed polar codes that utilize an upper triangular matrix for pre-transformation, our resulting pre-transformation matrix across multiple layers can be seen as a sub-matrix of an upper triangular matrix. For instance, we express our two-layered encoding structure with a unified pre-transform matrix ${\bf T}\in \mathbb{F}_2^{32\times 32}$ as follows:
\begin{align}
	[{\bf u}_{\mathcal{F}_2}, {\bf u}_{1}, {\bf u}_{\mathcal{I}_2}]\underbrace{\begin{bmatrix}
{\bf 0}_{17} & {\bf 0}_{17} & {\bf 0}_{17} \\
{\bf 0}_8 &  {\bf G}_8^{\top}& {\bf 0}_8  \\
{\bf 0}_{7} & {\bf 0}_{7} & {\bf I}_{7}   
\end{bmatrix}}_{{\bf T}}{\bf G}_{32}={\bf x}.
\end{align}
It is evident that the resulting pre-transformed matrix ${\bf T}$ does not exhibit an upper diagonal structure; instead, a sub-matrix of ${\bf T}$ takes the form of an upper triangular matrix. This condition is less strict than the PAC and the existing pre-transformed polar codes, where the entire ${\bf T}$ matrix must be  Toeplitz or upper triangular.

\section{deep polar Decoders}
In this section, we present two decoding algorithms for deep polar codes. The first decoding method is SCL with backpropagation parity check (SCL-BPC), which provides flexibility in balancing complexity and performance through list size control. Motivated by the superposition code interpretation, the second algorithm is the parallel-SCL, which aims to reduce the decoding latency at the cost of increased hardware complexity. 


\subsection{SCL-BPC Decoder}

 The central concept behind SCL-BPC is to efficiently prune decoding paths that fail to satisfy the successive parity check equations within the deep polar encoder during the SCL decoding process.  As a stepping stone toward understanding the overall SCL-BPC decoder, it is instructive to explain the BPC mechanism used in decoding, which is the key ingredient of the proposed decoder. The key concept behind the BPC mechanism is to leverage the reverse process of the deep polar encoder. From \eqref{eq:encode}, the deep polar encoder generates the connection bits of layer $\ell\in [L]$, ${\bf u}_{\mathcal{A}_{\ell}}$, using the input of encoder at layer $\ell-1$ and ${\bf G}_{N_{\ell-1}}^{\top}$ as 
\begin{align}
	{\bf u}_{\mathcal{A}_{\ell}}=[  {\bf u}_{\mathcal{I}_{\ell-1}},~ {\bf u}_{\mathcal{A}_{\ell-1}},~{\bf u}_{\mathcal{F}_{\ell-1}} ] {\bf G}_{N_{\ell-1}}^{\top}.
\end{align} 
 From this successive encoder structure, we know that if ${\bf u}_{\mathcal{A}_{\ell}}$ is successfully decoded, the reverse encoding using ${\bf T}_{\ell-1}^{-1}={\bf G}_{N_{\ell-1}}^{\top}$ ensures to produce the frozen bits of the previous layer:
 \begin{align}
 	{\bf u}_{\mathcal{F}_{\ell-1}} =	\left({\bf u}_{\mathcal{A}_{\ell}}{\bf G}_{N_{\ell-1}}^{\top} \right)_{{\mathcal{F}_{\ell-1}}},
 \end{align}
 for $\ell\in \{2,3,\ldots, L\}$. Precisely estimating ${\bf \hat u}_{\mathcal{A}_{L}}$ is crucial for effective decoding using the BPC mechanism. However, accurately estimating the connection bits in layer $L$ presents a significant challenge due to their transmission over less reliable bit-channels than the information bits during SCL decoding. Consequently, incorrect estimation of ${\bf \hat u}_{\mathcal{A}_{L}}$ can lead to a degradation in the performance of SCL decoding, mainly when the list size is insufficiently large.

\begin{figure*}[t]
\centering
\includegraphics[width=1.6\columnwidth]{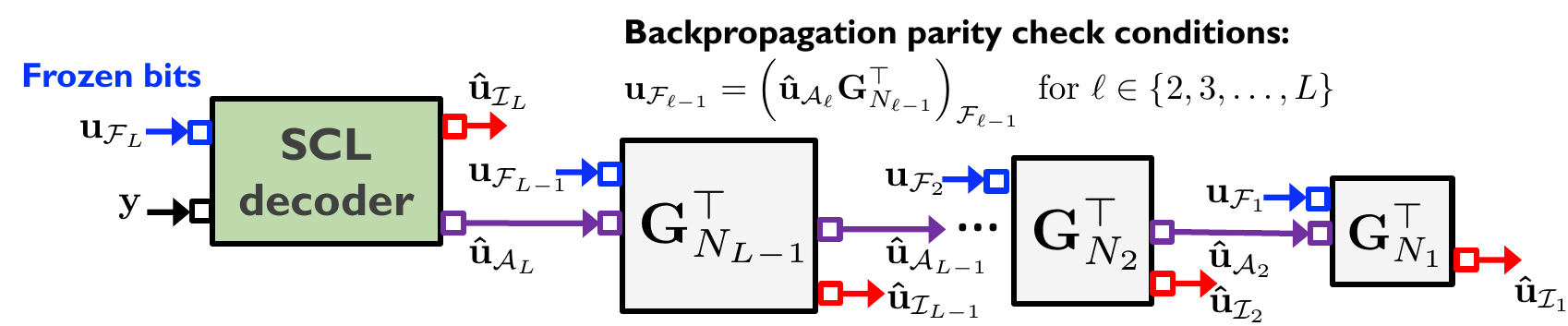}
\caption{An illustration of the SCL decoder using the backpropagation parity check principle.}
\label{fig:decoder}
\end{figure*}

\begin{algorithm}
\caption{ SCL-BPC Decoder}\label{alg:deep polar-SCL}
\KwData{List size $S$, channel output ${\bf y}$, transform matrices ${\bf T}_{\ell}$, information sets $\mathcal{I}_{\ell}$, connection sets $\mathcal{A}_{\ell}$, frozen sets $\mathcal{F}_{\ell}, ~ \forall \ell\in [L]$, and frozen bits ${\bf u}_{\mathcal{F}_L}$.}
\KwResult{Estimated information bits ${\bf \hat u}_{\mathcal{I}_{\ell}}[s^\star], ~ \forall \ell \in [L]$\;}
\vspace{0.2cm}
\emph{*** Initialization ***}\;
$\mathcal{P}_{\sf alive} \gets \{1\}$;
$\mathcal{P}_{\sf pool} \gets [2S]/\mathcal{L}_{\sf alive}$\; 
${\sf PM}[s] \gets 0, ~\forall s\in [S]$\;
\vspace{0.2cm}
\emph{*** Decoding ***}\;
\For(){$i=1,2,\ldots, N_L$}{
    \For(\tcp*[h]{path cloning}){$s\in \mathcal{P}_{\sf alive}$} {
        $\eta_i \gets \log \frac{ p\left({\bf y}, {\bf \hat u}_{L,1:i-1}[s] \vert u_{L,i} = 0\right) } { p\left({\bf y}, {\bf \hat u}_{L,1:i-1}[s] \vert u_{L,i} = 1\right) }$\;
        \eIf(\tcp*[h]{frozen bits}){$i\in \mathcal{F}_{L}$}{
            ${\hat u}_{L,i}[s] \gets u_{L,i}$\;
            ${\sf PM}[s] \gets {\sf PM}[s] + \log\left(1 + e^{-(1-2{\hat u}_{L,i}[s])\eta_i}\right)$\;}
            (\tcp*[h]{information bits}){
            Copy the path $s$ into a new path $s'\in \mathcal{P}_{\sf pool}$\;
            $\mathcal{P}_{\sf alive}\gets \mathcal{P}_{\sf alive} \cup \{s'\}$;
            $\mathcal{P}_{\sf pool} \gets \mathcal{P}_{\sf pool} / \{s'\}$\;
            ${\hat u}_{L,i}[s] \gets 0$; ${\hat u}_{L,i}[s'] \gets 1$\; 
            ${\sf PM}[s] \gets {\sf PM}[s] + \log\left(1 + e^{-(1-2{\hat u}_{L,i}[s])\eta_i}\right)$\;
            ${\sf PM}[s'] \gets {\sf PM}[s'] + \log\left(1 + e^{-(1-2{\hat u}_{L,i}[s'])\eta_i}\right)$\;          
        }
    }
    
    \If(\tcp*[h]{parity check}) {$i \in \mathcal{A}_{L}$}{
        \For {$s\in\mathcal{P}_{\sf alive}$} {
            Apply inverse transform according to \eqref{eqn:partial-inverse}\;
            ${\bf \hat u}_{\mathcal{F}_{\ell}, 1:k_{\ell}} \gets $ currently available frozen bits of layer $\ell\in[L]$\;
            \If {${\bf \hat u}_{\mathcal{F}_{\ell}, 1:k_{\ell}} \neq {\bf 0}$ for some $\ell$} {
                Kill the path $s$ \;
                $\mathcal{P}_{\sf alive}\gets \mathcal{P}_{\sf alive} /\{s\}$; $\mathcal{P}_{\sf pool} \gets \mathcal{P}_{\sf pool} \cup \{ s \}$ \;
            }
        }
    }    
    
    \If(\tcp*[h]{list pruning}) {$|\mathcal{P}_{\sf alive}| > S$} {
        $\tau_{\sf threshold} \gets$ the $S$th smallest ${\sf PM}[s]$ for $s\in\mathcal{P}_{\sf alive}$\;
        \For{${s}\in\mathcal{P}_{\sf alive}$}{
            \If {${\sf PM}[s] > \tau_{\sf threshold}$} {
                Kill the path $s$\;
                $\mathcal{P}_{\sf alive}\gets \mathcal{P}_{\sf alive} /\{s\}$; $\mathcal{P}_{\sf pool} \gets \mathcal{P}_{\sf pool} \cup \{ s \}$ \;
            }
        }
    }
}
Find $s^{\star}\gets \arg \min_{s\in \mathcal{P}_{\sf alive}} {\sf PM}[s]$\;
\vspace{0.2cm}
\emph{*** Information bits extraction ***}\;
\For{$\ell = L, \ldots, 2$}{
    ${\bf \hat u}_{\mathcal{I}_{\ell-1}}[s^\star] \gets \left( {\bf \hat u}_{\mathcal{A}_{\ell}}[s^\star] {\bf G}_{N_{\ell-1}}^{\top} \right)_{\mathcal{I}_{\ell-1}}$\;
    ${\bf \hat u}_{\mathcal{A}_{\ell-1}}[s^\star] \gets \left( {\bf \hat u}_{\mathcal{A}_{\ell}}[s^\star] {\bf G}_{N_{\ell-1}}^{\top} \right)_{\mathcal{A}_{\ell-1}}$\;
}
Return ${\bf \hat u}_{\mathcal{I}_{\ell}}[s^\star], ~ \forall \ell \in [L]$\;
\end{algorithm}

We propose an efficient SCL-BPC decoder by leveraging a bit-wise BPC mechanism to improve the decoding performance. The major idea is to verify the backpropagation syndrome check condition at the bit level within each layer, specifically for the elements denoted as $u_{L,i}$, where $i$ belongs to the set $\mathcal{A}_L$. To simplify our notation, we define ${\bf G}_{N_{\ell}, 1:k}^{\top} \in \mathbb{F}_2^{k\times k}$ as the upper-left submatrix of ${\bf G}_{N_{\ell}}^{\top}$. It is important to note that our pre-transform technique using the polar transform matrix possesses a unique property: its inverse is equal to itself. As a result, the inverse pre-transformation matrix is simply calculated itself as 
\begin{align}
    \left({\bf G}_{N_{\ell}, 1:k}^{\top}\right)^{-1} = {\bf G}_{N_{\ell}, 1:k}^{\top}, ~\forall \ell\in[L-1], \forall k\in[K].
\end{align}
If we denote the connection bits of layer $\ell$ as ${\bf \hat u}_{\mathcal{A}_{\ell}} = [{\hat u}_{\ell, i_{1}}, {\hat u}_{\ell, i_{2}}, \ldots, {\hat u}_{\ell, i_{N_{\ell-1}}}]$ such that $i_1 < i_2 < \cdots < i_{N_{\ell-1}}$, we express 
\begin{align}
    {\bf \hat u}_{\ell-1, 1:k} = {\bf \hat u}_{\mathcal{A}_{\ell}, 1:k} {\bf G}_{N_{\ell-1}, 1:k}^{\top}, \label{eqn:partial-inverse}
\end{align}
where ${\bf \hat u}_{\ell-1, 1:k}$ is a subsequence that comprises the first $k$ elements of ${\bf \hat u}_{\ell-1}$. Next, we extract two portions from the estimated bits ${\bf \hat u}_{\ell-1}$ of layer $\ell-1$: one portion is used for parity checking, and the other portion consists of connection bits that are recursively applied using  \eqref{eqn:partial-inverse}. Finally, we gather the estimated frozen bits from each layer $\ell \in [L]$ and verify their syndromes as depicted in Fig.~\ref{fig:decoder}.

Leveraging this bit-wise BPC mechanism, the proposed SCL decoder employs a level-by-level search strategy on a binary tree. At each bit $u_{L,i}$, where $i\in \mathcal{A}_L\cup \mathcal{I}_L$, the decoder extends the list of candidate paths by exploring both paths of the binary tree and appending either $u_{L,i}=0$ or $u_{L,i}=1$ to each candidate path. Consequently, the number of paths is doubled but limited to a predetermined maximum value $S$. In contrast to the standard SCL decoder, it is included in the list whenever a new path satisfies the BPC and ranks among the $S$ most reliable paths. Conversely, it is discarded if a new path fails the BPC or exhibits lower reliability than the existing $S$ paths in the list. This iterative process continues until $i\in [N_L]$. Ultimately, the decoder selects the path with the highest reliability metric as the output. When $S=1$, our decoder simplifies to the SC decoder utilizing the backpropagation parity check. The deep polar SCL decoding procedure is summarized in Algorithm~\ref{alg:deep polar-SCL}.
  
    The proposed decoder introduces an extra decoding complexity due to the BPC operation compared to the standard SCL decoding complexity with a list size of $S$, which is $\mathcal{O}(SN_L\log N_L)$. For each layer $\ell\in \{1,2,\ldots, N_{L-1}\}$, the parity check can be performed with a complexity of $\mathcal{O}(N_{\ell}\log N_{\ell})$. Consequently, the overall complexity is the sum of the complexity required for SCL decoding and computing the inverse of the pre-transforms, i.e.,
\begin{align}
	\mathcal{O}(SN_L\log N_L)+ \mathcal{O}\left(\sum_{\ell=1}^{L-1}N_{\ell}\log N_{\ell}\right).
\end{align} 
It is worth noting that the additional complexity introduced by the inverse of the pre-transform can be ignored when $N_L$ is significantly larger than $N_{\ell}$ for $\ell\in[L-1]$.

  \subsection{A Low-Latency Decoder }
  
   \begin{figure*}[t]
 \centering
 \includegraphics[width=1.7\columnwidth]{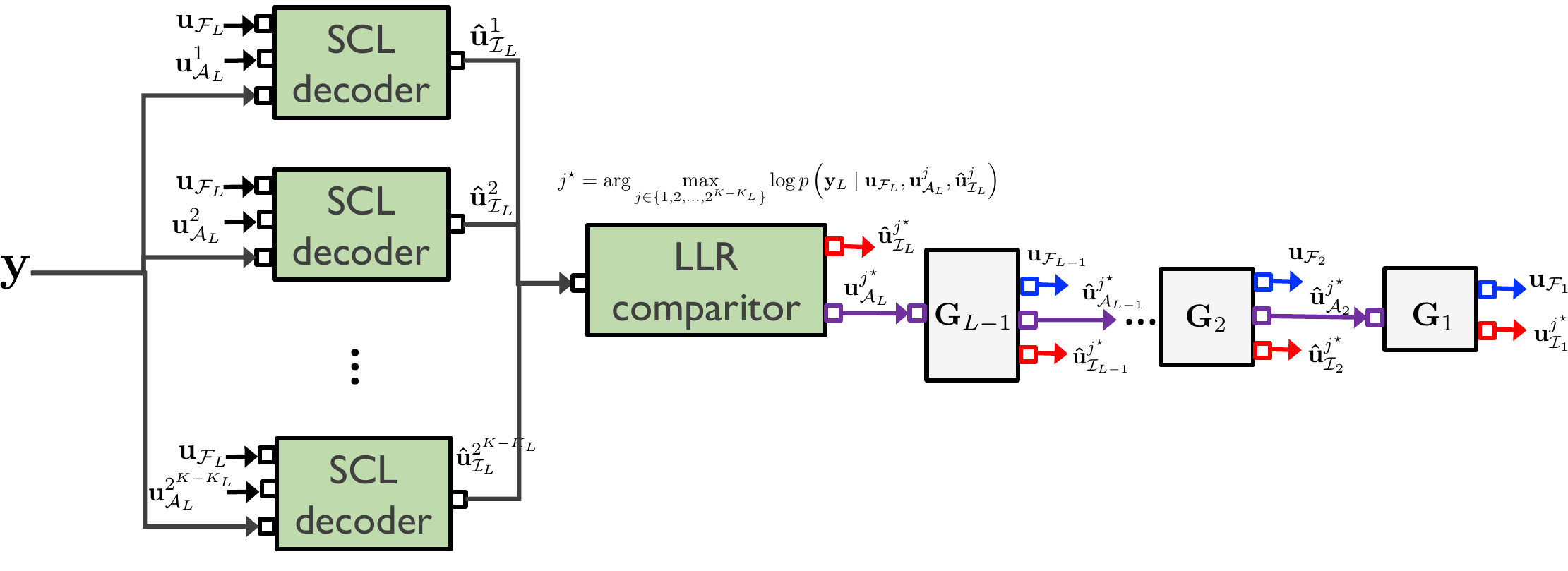}
 \caption{The proposed backpropagation decoding method using parallel-SCL decoders.}
 \label{fig:decoder-parallel}
 \end{figure*}

We also present a method for achieving low-latency decoding of deep polar codes. Our approach involves leveraging standard SCL decoders in parallel to estimate the information vector ${\bf u}_{\mathcal{I}_{L}}$, assuming the connection bits in layer $L$, ${\bf u}_{\mathcal{A}_{L}}$, as an additional frozen set alongside with frozen bits ${\bf u}_{\mathcal{F}_L}$. There are $2^{K-K_L}$ potential connection bit patterns for ${\bf u}_{\mathcal{A}_{L}}$. Let ${\bf u}_{\mathcal{A}_{L}}^j\in \mathbb{F}_2^{N_{L-1}}$ represent the $j$th connection bit pattern, where $j\in \left\{1,2,\ldots, 2^{K-K_L}\right\}$. Given the $j$th connection bit pattern ${\bf u}_{\mathcal{A}_{L}}^j$ and the frozen bits of the last layer ${\bf u}_{\mathcal{F}_{L}}={\bf 0}$, the SCL decoder identifies the most reliable information vector from a list of size $S$. Denoting the result of SCL decoding for the $j$th possible connection bit pattern and frozen bits as ${\bf \hat u}_{\mathcal{I}_{L}}^{j}$, the decoder selects the most reliable estimate ${\bf \hat u}_{\mathcal{I}_{L}}^{j^{\star}}$ for the information vector from the set $\{1,2,\ldots, 2^{K-K_L}\}$ as 
\begin{align}
 j^{\star}=\argmax_{j\in \{1,2,\ldots, 2^{K-K_L}\} }  \log p\left({\bf y}_L \mid {\bf u}_{\mathcal{F}_{L}}, {\bf u}_{\mathcal{A}_{L}}^j, {\bf \hat u}_{\mathcal{I}_{L}}^j\right).
\end{align}
The parallel-SCL decoding method diminishes the decoding latency by harnessing the power of parallel processing. However, the hardware complexity of this method increases exponentially with the number of information bits encoded in $\ell$ layers, where $\ell\in \{1,2,\ldots, L-1\}$ and $K-K_L$. Consequently, practical usage of this method is limited to scenarios where the number of $K-K_L$ is small enough. In addition, the overall computational complexity of the parallel-SCL decoding becomes 
\begin{align}
	\mathcal{O}\left(2^{K-K_L}SN_L\log N_L\right).
\end{align} 
The list size $S$ of this parallel-SCL decoding can be chosen very small because the information bits are sent through the bit-channels with the highest capacities under the premise that the actual connection bits are utilized as the frozen bits. Therefore, the implementation needs to carefully optimize the list size and the number of parallel-SCL decoders so that $2^{K-K_L}S$ is comparable to the other SCL-type decoders.  

\section{Simulation Results}
This section compares the BLERs of different encoding and decoding methods under BI-AWGN channel.

\begin{table*}
\centering
\caption{Simulation parameters} \label{table:simulation-parameters}
\begin{tabular}{llllllBBBBBB}
\toprule
{} & {} & {} & {} & {} & {} & \multicolumn{2}{c}{${\sf d}_{L}^{\rm min}$} & \multicolumn{4}{c}{$(N_{\ell}, K_{\ell})$} \\
\cmidrule(lr){7-8}\cmidrule(lr){9-12}
{} & Fig. & {$(N,K)$} & {Rate-profile} & {Pre-transform} & {Decoder} & {Design} & {Estimate} & {$\ell = 1$} & {$\ell = 2$} & {$\ell = 3$} & {$\ell=4$} \\
\midrule
{DP} & \ref{fig:BLER-PAC} &$(128,29)$ & 5G \cite{3gpp-nr-coding} & Polar & SCL-BPC & $16$ & $32$ & $(128,19)$ & $(16,8)$ & $(4,2)$ & {} \\
{DP} & \ref{fig:BLER-PAC} (---) &$(128,64)$ & 5G \cite{3gpp-nr-coding} & Polar & SCL-BPC & $8$ & $8$ & $(128,51)$ & $(16,13)$ & {} & {} \\
{CA-DP} & \ref{fig:BLER-PAC} (-\,-) &$(128,64)$ & 5G \cite{3gpp-nr-coding} & Polar + CRC6 & SCL-BPC & $8$ & $12$ & $(128,67)$ & $(16,3)$ & {} & {} \\
\midrule
{DP} & \ref{fig:BLER-CA-Polar} & $(128,32)$ & 5G \cite{3gpp-nr-coding} & Polar & SCL-BPC & $16$ & $24$ & $(128,21)$ & $(16,6)$ & $(8,3)$ & $(4,2)$ \\
{DP} & \ref{fig:BLER-CA-Polar} &$(128,56)$ & 5G \cite{3gpp-nr-coding} & Polar & SCL-BPC & $16$ & $16$ & $(128,52)$ & $(8,4)$ & {} & {} \\
{DP} & \ref{fig:BLER-CA-Polar} &$(128,96)$ & 5G \cite{3gpp-nr-coding} & Polar & SCL-BPC & $8$ & $8$ & $(128,94)$ & $(4,2)$ & {} & {} \\
\midrule
{DP} & \ref{fig:BLER-PAC-parallel}-\ref{fig:BLER-DP-vs-SCL} &$(128,29)$ & DEGA 1.5 dB & Polar & parallel-SCL & 16 & 32 & $(128,24)$ & $(32,3)$ & $(8,1)$ & $(2,1)$ \\
{DP} & \ref{fig:BLER-PAC-parallel}-\ref{fig:BLER-DP-vs-SCL} &$(128,64)$ & DEGA 6 dB & Polar & parallel-SCL & 8 & 16 & $(128,59)$ & $(32,3)$ & $(8,1)$ & $(2,1)$ \\
\midrule
{DP} & \ref{fig:BLER-BOSS} &$(64,16)$ & 5G \cite{3gpp-nr-coding} & Polar & ML & $16$ & $16$ & $(64,6)$ & $(16,8)$ & $(4,2)$ & {} \\
{DP} & \ref{fig:BLER-BOSS} &$(128,16)$ & 5G \cite{3gpp-nr-coding} & Polar & ML & $32$ & $32$ & $(128,6)$ & $(16,8)$ & $(4,2)$ & {} \\
{DP} & \ref{fig:BLER-BOSS} &$(256,16)$ & 5G \cite{3gpp-nr-coding} & Polar & ML & $64$ & $64$ & $(256,6)$ & $(16,8)$ & $(4,2)$ & {} \\
{CA-DP} & \ref{fig:BLER-BOSS} &$(64,16)$ & 5G \cite{3gpp-nr-coding} & Polar + CRC6 & ML & $8$ & $16$ & $(64,11)$ & $(16,11)$ & {} & {} \\
{CA-DP} & \ref{fig:BLER-BOSS} &$(128,16)$ & 5G \cite{3gpp-nr-coding} & Polar + CRC6 & ML & $32$ & $32$ & $(128,12)$ & $(16,10)$ & {} & {} \\
{CA-DP} & \ref{fig:BLER-BOSS} &$(256,16)$ & 5G \cite{3gpp-nr-coding} & Polar + CRC6 & ML & $64$ & $96$ & $(256,13)$ & $(16,9)$ & {} & {} \\
\midrule
CA-polar & \ref{fig:BLER-PAC}-\ref{fig:BLER-BOSS} &$\forall (N,K)$ & 5G \cite{3gpp-nr-coding} & CRC6 & SCL & {} & {} & {} & {} & {} & {} \\
\midrule
PAC & \ref{fig:BLER-PAC},\ref{fig:BLER-PAC-parallel} &$(128,29)$ & RM \cite{arikan-pac} & CC $({\bf c} = 3211)$ & SCL & {} & {$32$} & {} & {} & {} & {} \\
PAC & \ref{fig:BLER-PAC},\ref{fig:BLER-PAC-parallel} &$(128,64)$ & RM \cite{arikan-pac} & CC $({\bf c} = 133)$ & SCL & {} & {$16$} & {} & {} & {} & {} \\
\bottomrule
\multicolumn{12}{l}{** DP: deep polar,~ DEGA: Density Evolution with Gaussian Approximation \cite{Trifnov-polar-construction},~ CC: Convolutional Coding } \\
\multicolumn{12}{l}{** Estimate ${\sf d}_{L}^{\rm min}$ is computed using SCL(-BPC) decoder with list size $S=10^4$. } \\
\end{tabular}
\end{table*}



\subsection{Proposed Code Construction}
We implement two types of deep polar codes with/without CRC precoding.
\begin{itemize}

	\item {\bf Deep polar codes:}  	 We have developed deep polar codes with varying numbers of layers, $L \in \{2, 3, 4\}$, based on the size of the information bits $K$. For instance, Table~\ref{table:simulation-parameters} shows encoding network configurations for deep polar codes. A significant finding from our investigations is that the minimum distance of the deep polar code, $d^{\sf min}$, can exceed the minimum weight of the selected rows from ${\bf G}_{128}$ in $\mathcal{A}_L$, i.e., $d_{L}^{\sf min}$. This observation indicates that using multi-layered pre-transformations can increase the minimum distance of the code.

	 	    \item {\bf CA-deep polar codes:} We employ CRC precoding as an extra pre-transformation before applying the successive deep polar encoding. We utilize the specific 6-bit CRC polynomial $1 + D^5 + D^6$ for the CRC precoding. The effectiveness of CRC precoding in enhancing the codes is demonstrated in Table~\ref{table:simulation-parameters}, where it is observed that the minimum distance is increased with only a small number of layers for encoding. For example, the $(128,64)$ CA-deep polar code with two layers exhibits an increased minimum distance compared to the two-layered deep polar code without CRC precoding. This result suggests that CRC precoding serves as an additional pre-transformation that introduces more layers into the encoding process. Consequently, it contributes to the overall improvement in the performance of the deep polar codes.

	    \item {\bf Decoders:} We employ three decoding algorithms for the deep polar codes. These algorithms include the proposed SCL-BPC decoding and parallel-SCL decoding techniques with a list size of $S$, which are primarily utilized to reduce decoding complexity and delays during simulations. The ML decoding algorithm is also employed to gauge the gap-to-capacity performance specifically for low code rates, such as $K=16$ and $N\in \{64, 128, 256\}$.

\end{itemize}

\subsection{Benchmarks}

We explain benchmark encoding and decoding schemes to compare the BLER performance in a short blocklength regime. The benchmarks  are listed as follows:
\begin{itemize}
\item {\bf Theoretical bounds \cite{polyanskiy-finite}:}  We utilize theoretical benchmarks to assess the performance of our proposed encoding and decoding methods. Specifically, we plot the dispersion normal approximation and the meta-converse bounds for a given code length and rates, as described in \cite{polyanskiy-finite}. The dispersion and meta-converse bounds were calculated using the {\it spectre} package, which can be accessed online at \url{https://github.com/yp-mit/spectre}. These bounds allow us to demonstrate the gaps between the achievable performance of our methods and the upper bounds of the finite-blocklength capacity.

	\item {\bf CA-polar codes \cite{3gpp-nr-coding}:} The CA-polar codes serve as the baseline methods. The information sets are chosen according to the 5G standard specification in \cite{3gpp-nr-coding} when constructing the polar codes. In addition, we take into account the CRC precoding with the polynomial of $1 + D^5 + D^6$, which has shown the lowest BLER among CRC polynomials in 3GPP standards \cite{3gpp-nr-coding}. For decoding the CA-polar codes, we use the standard SCL decoder with a list size of $S$.
	
	\item {\bf PAC codes  \cite{arikan-pac}:} The PAC codes heavily depend on the rate-profile algorithm and generator polynomials of convolutional code. The optimized rate-profiling method is known for information bit sizes of $K=29$ and $K=64$ under blocklength of 128 \cite{arikan-pac}. Specifically, the PAC codes adopt the generator polynomials ${\bf c} = 3211$ and ${\bf c} = 133$ in the octal notation for $K=29$ and $K=64$, respectively.
	 
	\item {\bf CA-BOSS codes \cite{BOSS-URLLC}:} We also consider comparing a block orthogonal sparse superposition (BOSS) code. Concatenating with CRC codes, the CA-BOSS code was shown to achieve the meta converse bounds within 1 dB for a short blocklength regime. For comparison, we implement a CA-BOSS code with the CRC polynomial of $1 + D + D^3$. We refer the reader to \cite{BOSS-URLLC} for the construction details.
\end{itemize}

\subsection{BLER Comparisons}

\begin{figure}[t]
\centering
\includegraphics[width=1.1\columnwidth]{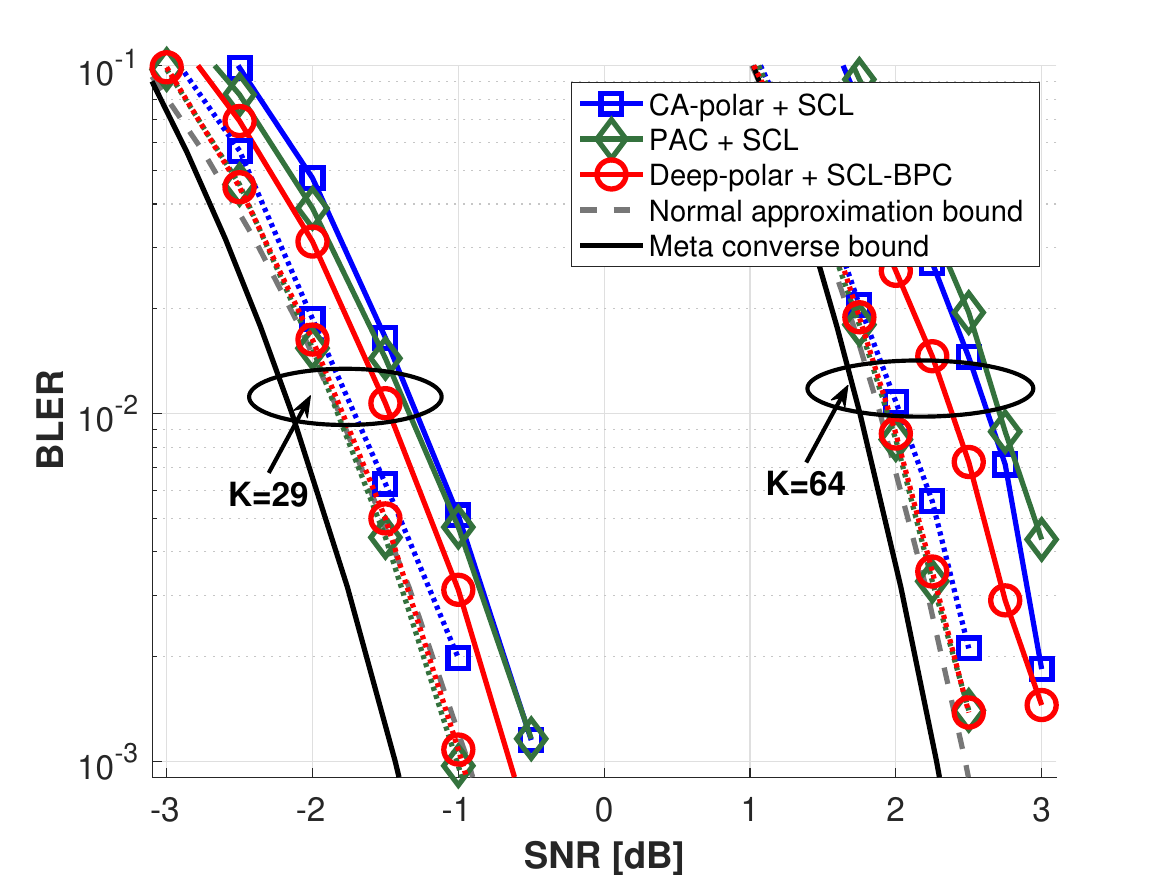}
\caption{BLER performance of PAC, CA-polar, and deep polar codes. Parameters for solid line are $(K,S) \in \{(29,8), (64,8)\}$, and for dotted line are $(K,S)\in \{(29,32), (64,256)\}$.}
\label{fig:BLER-PAC}
\end{figure}

{\bf Comparison with PAC codes:} Fig.~\ref{fig:BLER-PAC} presents the comparisons of BLER performances among three coding schemes: deep polar codes, the PAC codes with optimized rate-profiling, and 5G standard CA-polar codes, each evaluated at two distinct code rates, $R \in \{\frac{29}{128},\frac{64}{128}\}$. The meta-converse and the normal approximation bounds are plotted as theoretical benchmarks for the corresponding code rates. For decoding, SCL-BPC decoders are utilized for deep polar codes, while SCL decoders are employed for both PAC and CA-polar codes with varying list sizes, $S \in \{8,32,256\}$. The solid lines on the graph correspond to the BLER performances of SCL-BPC and SCL decoders when using a short list size of $S=8$. The results demonstrate that deep polar codes outperform PAC and CA-polar codes when employing a short list size, particularly where $R=\frac{64}{128}$, PAC codes exhibit a worse BLER performance than CA-polar codes. Additionally, we observe the dotted lines on the plot, representing the BLER performances with larger list sizes, i.e., $S=32$ for $K=29$ and $S=256$ for $K=64$. Notably, with sufficiently large list sizes, both deep polar and PAC codes closely achieve the normal approximation bound, while CA-polar codes still exhibit a gap from the bound.




 \vspace{0.2cm}

\begin{figure}[t]
\centering
\includegraphics[width=1.1\columnwidth]{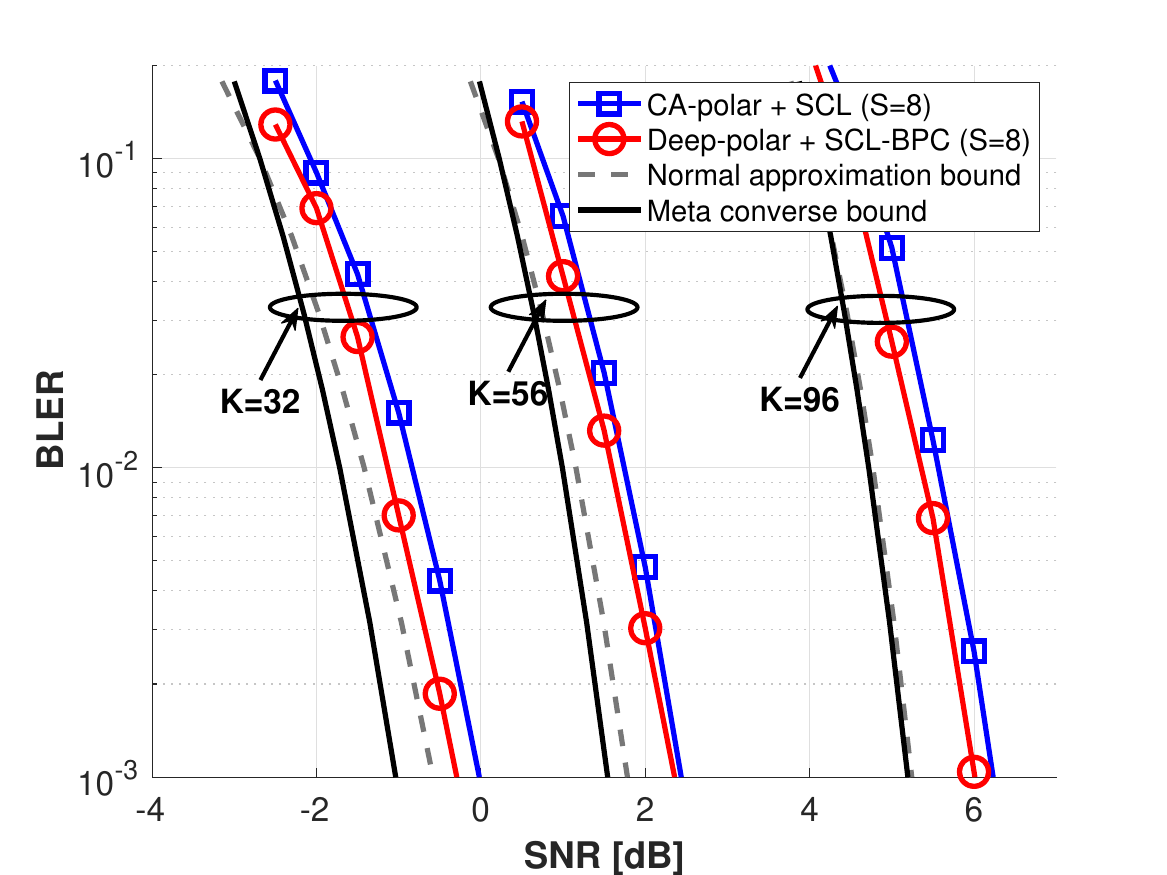}
\caption{BLER performance comparison bewteen deep polar and CA-polar codes. The code parameters are $N=128$ and $K\in\{32, 56, 96$\}.}
\label{fig:BLER-CA-Polar}
\end{figure}

{\bf Comparison with CA-polar codes:}
Fig.~\ref{fig:BLER-CA-Polar} illustrates the comparison of BLER performance between the proposed deep polar and 5G CA-polar codes using two decoding methods: SCL-BPC and SCL with a list size of $S=8$. The simulation result reveals that our deep polar codes consistently outperform the CA-polar counterparts at three specific code rates, namely $R\in \{\frac{32}{128}, \frac{56}{128}, \frac{96}{128}\}$. This result suggests that the proposed deep polar code exhibits superior performance compared to the 5G CA-polar codes across various code rates, while the increase in decoding complexity remains marginal.

\begin{figure}[t]
\centering
\includegraphics[width=1.1\columnwidth]{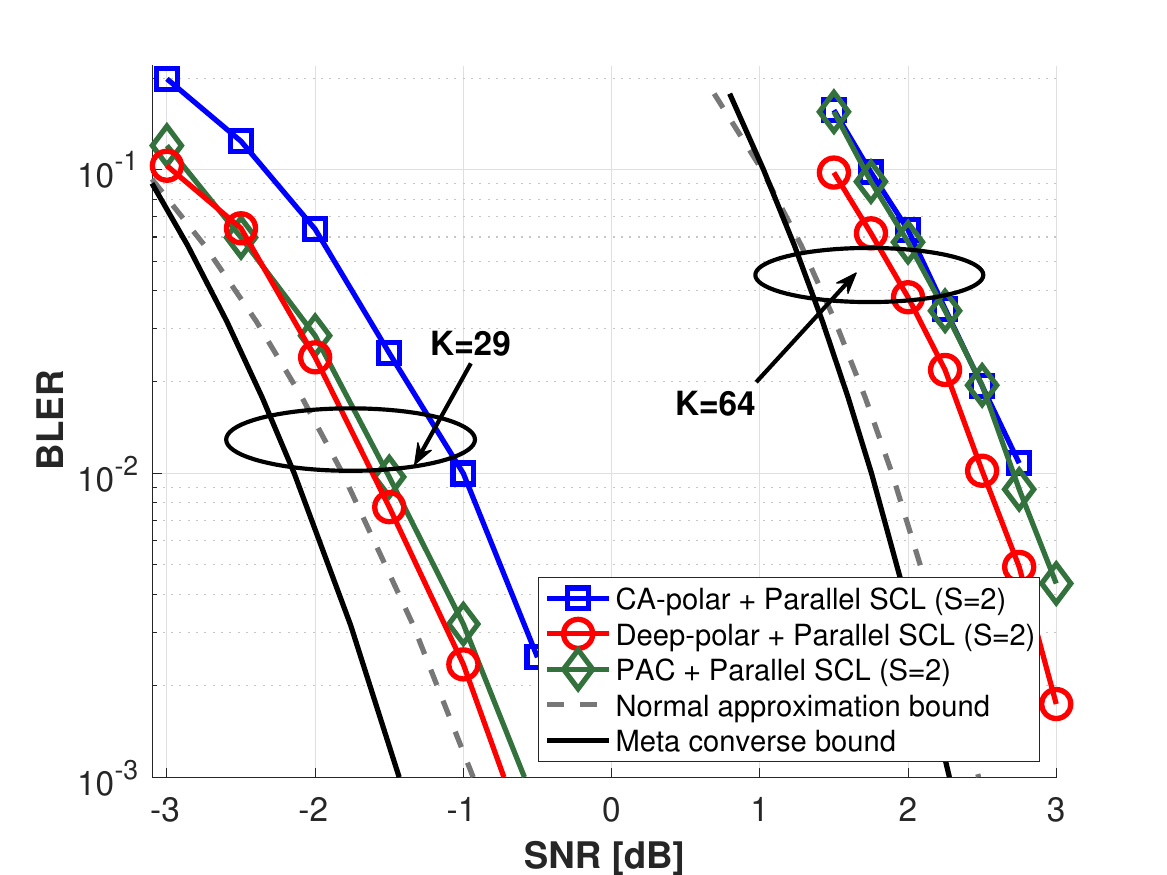}
\caption{BLER performance under parallel-SCL decoder.}
\label{fig:BLER-PAC-parallel}
\end{figure}

 \vspace{0.2cm}
{\bf Performance of parallel-SCL decoder:}
 Fig.~\ref{fig:BLER-PAC-parallel} shows the BLER performance for the deep polar, PAC, and CA-polar codes under parallel-SCL decoding. To implement the parallel-SCL decoder for the PAC and CA-polar codes, we select the $K-K_L$ smallest information indices and generate $2^{K-K_L}$ combinations of information bit patterns as possible frozen bit patterns. Then, we apply the parallel-SCL decoder by adding the generated information bit patterns as additional frozen bits. For each possible frozen bit pattern, the decoder applies the SCL decoding with a list size of $S=2$. The simulation result shows that deep polar codes outperform CA-polar and PAC codes under parallel-SCL decoding by attesting that local pre-transform gives robustness to parallel-SCL decoding.

\begin{figure}[t]
\centering
\includegraphics[width=1.1\columnwidth]{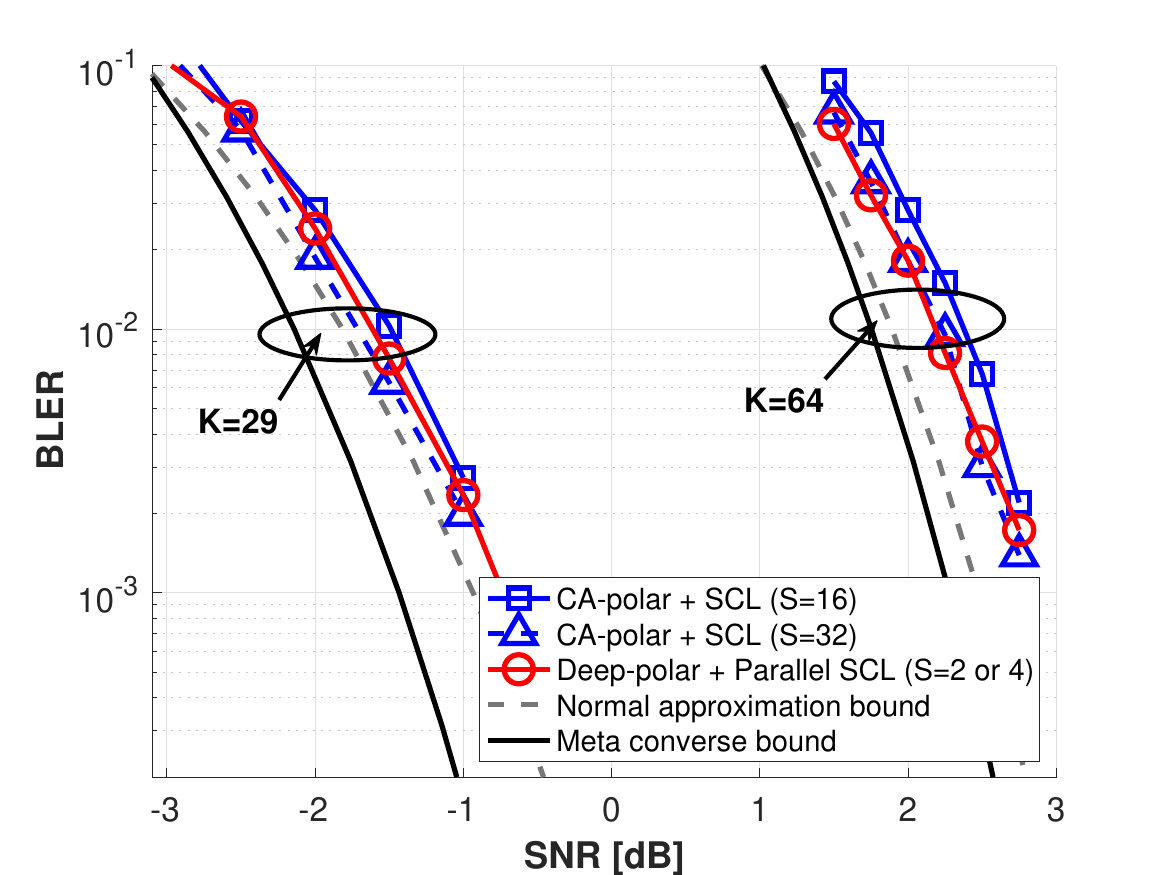}
\caption{BLER performance of deep polar codes under parallel-SCL decoder and CA-polar codes under SCL decoder. We use list size $S=2$ for decoding deep polar codes for $K=29$, and $S=4$ for $K=64$.}
\label{fig:BLER-DP-vs-SCL}
\end{figure}

Fig.~\ref{fig:BLER-DP-vs-SCL} shows the BLER performance of deep polar codes decoded by parallel-SCL and CA-polar codes by SCL decoder. We use a parallel-SCL decoder with list size $S=2$ to decode deep polar codes with message length $K=29$ and list size $S=4$ for $K=64$. The BLER curve of deep polar codes under parallel-SCL decoder lies between that of CA-polar codes under SCL decoder with list size $S=16$ and $S=32$. Even if the total number of computations of parallel-SCL is larger than SCL counterparts, we can parallelize the decoding operation to achieve low decoding latency, thereby attesting to the suitability of deep polar codes for low-latency communication applications.

\begin{figure}[t]
\centering
\includegraphics[width=1.1\columnwidth]{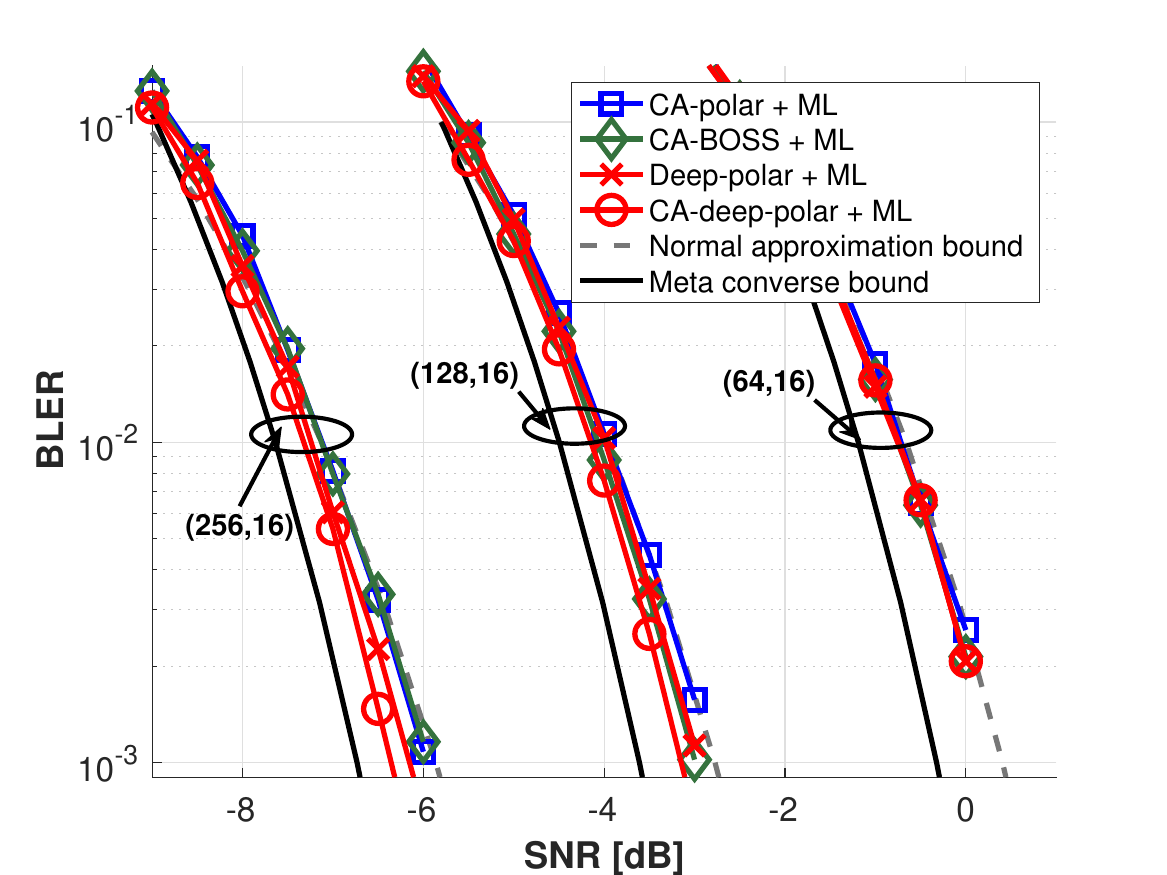}
\caption{BLER performance of deep polar, CA-deep polar, CA-polar, and CA-BOSS codes. The CRC polynomial is $1 + D^5 + D^6$ for CA-deep polar and CA-polar codes and $1 + D + D^3$ for CA-BOSS codes.}
\label{fig:BLER-BOSS}
\end{figure}

{\bf Comparison with CA-BOSS codes:}
Fig.~\ref{fig:BLER-BOSS} presents the ML decoding performance of four coding schemes: deep polar codes, CA-deep polar codes, CA-BOSS codes, and CA-polar codes, at different code rates $R\in \{\frac{16}{256},\frac{16}{128},\frac{16}{64}\}$. The results demonstrate the effectiveness of the proposed codes in improving the weight spectrum. By utilizing ML decoding, we assess the performance improvement achieved by the proposed codes. For example, when $R=\frac{16}{128}$, the CA-BOSS code marginally outperforms the deep polar code. However, with the integration of CRC precoding, the CA-deep polar code exhibits superior decoding performance compared to all other considered coding schemes, even achieving the meta converse bound within 0.4 dB. This remarkable outcome highlights the significant benefits of harmonizing local pre-transform (e.g., deep polar code) with global pre-transform (e.g., CRC) for boosting decoding performance at low rates.


\section{Conclusion}
 We have introduced a novel type of pre-transformed polar code called the deep polar code. The deep polar codes are generated through successive multi-layered encoding with a flexible rate-profiling method. The main technical innovation involves creating a unique yet universal pre-transformed architectures of polar codes, which is inspired by deep layered transform coding. These innovative architectures open up new possibilities for exploring a far more diverse range of pre-transformed polar code structures that were not explored before. Our deep polar code is a superposition code of pre-transformed and regular polar subcodes. Controlling the rates of these two subcodes enables the design of low-complexity decoding while significantly improving the weight distribution of the codes. Leveraging this superposition property, we have also introduced two decoding methods: the SCL-BPC decoder, which effectively reduces decoding complexity, and the parallel-SCL decoder, which minimizes decoding latency. Through simulations, we have demonstrated that our codes achieve close to finite-blocklength capacity and consistently outperform all existing state-of-the-art pre-transformed polar codes at various rates and short blocklengths, all while maintaining a reasonable decoding complexity.

A promising further research direction would be to extend the deep polar code design to moderate blocklengths, such as up to 2048, in order to bridge the gap towards the finite-blocklength capacity in moderate blocklength regimes. Additionally, exploring the development of a more efficient decoding algorithm for larger code blocklengths holds great promise, as the complexity of the SCL-BPC decoder may become computationally prohibitive in moderate blocklengths. Lastly, it is also interesting to design an efficient rate-compatible deep polar code to optimally support HARQ protocols.

\bibliographystyle{IEEEtran}
\bibliography{Coding}

\begin{thebibliography}{10}
\providecommand{\url}[1]{#1}
\csname url@samestyle\endcsname
\providecommand{\newblock}{\relax}
\providecommand{\bibinfo}[2]{#2}
\providecommand{\BIBentrySTDinterwordspacing}{\spaceskip=0pt\relax}
\providecommand{\BIBentryALTinterwordstretchfactor}{4}
\providecommand{\BIBentryALTinterwordspacing}{\spaceskip=\fontdimen2\font plus
\BIBentryALTinterwordstretchfactor\fontdimen3\font minus
  \fontdimen4\font\relax}
\providecommand{\BIBforeignlanguage}[2]{{%
\expandafter\ifx\csname l@#1\endcsname\relax
\typeout{** WARNING: IEEEtran.bst: No hyphenation pattern has been}%
\typeout{** loaded for the language `#1'. Using the pattern for}%
\typeout{** the default language instead.}%
\else
\language=\csname l@#1\endcsname
\fi
#2}}
\providecommand{\BIBdecl}{\relax}
\BIBdecl

\bibitem{Wang-6G-Survey}
C.-X. Wang, X.~You, X.~Gao, X.~Zhu, Z.~Li, C.~Zhang, H.~Wang, Y.~Huang,
  Y.~Chen, H.~Haas, J.~S. Thompson, E.~G. Larsson, M.~D. Renzo, W.~Tong,
  P.~Zhu, X.~Shen, H.~V. Poor, and L.~Hanzo, ``On the road to {6G}: Visions,
  requirements, key technologies, and testbeds,'' \emph{IEEE Commun. Surveys
  Tuts.}, vol.~25, no.~2, pp. 905--974, 2nd Quart. 2023.

\bibitem{tataria-6G-URLLC}
H.~Tataria, M.~Shafi, A.~F. Molisch, M.~Dohler, H.~Sj{\"o}land, and
  F.~Tufvesson, ``{6G} wireless systems: Vision, requirements, challenges,
  insights, and opportunities,'' \emph{Proceedings of the IEEE}, vol. 109,
  no.~7, pp. 1166--1199, Jul. 2021.

\bibitem{zhang-6G-URLLC}
Z.~Zhang, Y.~Xiao, Z.~Ma, M.~Xiao, Z.~Ding, X.~Lei, G.~K. Karagiannidis, and
  P.~Fan, ``{6G} wireless networks: Vision, requirements, architecture, and key
  technologies,'' \emph{IEEE Veh. Technol. Mag.}, vol.~14, no.~3, pp. 28--41,
  Sep. 2019.

\bibitem{david-6G-URLLC}
K.~David and H.~Berndt, ``{6G} vision and requirements: Is there any need for
  beyond {5G}?'' \emph{IEEE Veh. Technol. Mag.}, vol.~13, no.~3, pp. 72--80,
  Sep. 2018.

\bibitem{Chen-URLLC}
H.~Chen, R.~Abbas, P.~Cheng, M.~Shirvanimoghaddam, W.~Hardjawana, W.~Bao,
  Y.~Li, and B.~Vucetic, ``Ultra-reliable low latency cellular networks: Use
  cases, challenges and approaches,'' \emph{IEEE Commun. Mag.}, vol.~56,
  no.~12, pp. 119--125, Dec. 2018.

\bibitem{shirvanimoghaddam19}
M.~Shirvanimoghaddam, M.~S. Mohammadi, R.~Abbas, A.~Minja, C.~Yue, B.~Matuz,
  G.~Han, Z.~Lin, W.~Liu, Y.~Li, S.~Johnson, and B.~Vucetic, ``Short
  block-length codes for ultra-reliable low latency communications,''
  \emph{IEEE Commun. Mag.}, vol.~57, no.~2, pp. 130--137, Feb. 2019.

\bibitem{yue23}
C.~Yue, V.~Miloslavskaya, M.~Shirvanimoghaddam, B.~Vucetic, and Y.~Li,
  ``Efficient decoders for short block length codes in {6G URLLC},'' \emph{IEEE
  Commun. Mag.}, vol.~61, no.~4, pp. 84--90, Apr. 2023.

\bibitem{polyanskiy-finite}
Y.~Polyanskiy, H.~V. Poor, and S.~Verdu, ``Channel coding rate in the finite
  blocklength regime,'' \emph{IEEE Trans. Inf. Theory}, vol.~56, no.~5, pp.
  2307--2359, May 2010.

\bibitem{Yang-finite-mimo-fading}
W.~Yang, G.~Durisi, T.~Koch, and Y.~Polyanskiy, ``Quasi-static multiple-antenna
  fading channels at finite blocklength,'' \emph{IEEE Trans. Inf. Theory},
  vol.~60, no.~7, pp. 4232--4265, Jul. 2014.

\bibitem{Makki-finite-HARQ}
B.~Makki, T.~Svensson, and M.~Zorzi, ``Finite block-length analysis of the
  incremental redundancy {HARQ},'' \emph{IEEE Wireless Commun. Lett.}, vol.~3,
  no.~5, pp. 529--532, Oct. 2014.

\bibitem{arikan-polar}
E.~Arikan, ``Channel polarization: A method for constructing capacity-achieving
  codes for symmetric binary-input memoryless channels,'' \emph{IEEE Trans.
  Inf. Theory}, vol.~55, no.~7, pp. 3051--3073, Jul. 2009.

\bibitem{Hui-5G-coding-comparison}
D.~Hui, S.~Sandberg, Y.~Blankenship, M.~Andersson, and L.~Grosjean, ``Channel
  coding in {5G} new radio: A tutorial overview and performance comparison with
  {4G LTE},'' \emph{IEEE Veh. Technol. Mag.}, vol.~13, no.~4, pp. 60--69, Dec.
  2018.

\bibitem{Tal-polar-SCL}
I.~Tal and A.~Vardy, ``List decoding of polar codes,'' \emph{IEEE Trans. Inf.
  Theory}, vol.~61, no.~5, pp. 2213--2226, May 2015.

\bibitem{Balatsoukas-SCL-log}
A.~Balatsoukas-Stimming, M.~B. Parizi, and A.~Burg, ``{LLR}-based successive
  cancellation list decoding of polar codes,'' \emph{IEEE Trans. Signal
  Process.}, vol.~63, no.~19, pp. 5165--5179, Oct. 2015.

\bibitem{mondelli-urbanke-polar2RM}
M.~Mondelli, S.~H. Hassani, and R.~L. Urbanke, ``From polar to {Reed-Muller}
  codes: A technique to improve the finite-length performance,'' \emph{IEEE
  Trans. Commun.}, vol.~62, no.~9, pp. 3084--3091, Sep. 2014.

\bibitem{li-tse-rm-polar}
B.~Li, H.~Shen, and D.~Tse, ``A {RM}-polar codes,'' \emph{arXiv preprint
  arXiv:1407.5483}, 2014, [Online]. Available: https://arxiv.org/abs/1407.5483.

\bibitem{Abbe-Reed-Muller}
E.~Abbe, A.~Shpilka, and M.~Ye, ``{Reed–Muller} codes: Theory and
  algorithms,'' \emph{IEEE Trans. Inf. Theory}, vol.~67, no.~6, pp. 3251--3277,
  Jun. 2021.

\bibitem{Niu-CA-polar}
K.~Niu and K.~Chen, ``{CRC}-aided decoding of polar codes,'' \emph{IEEE Commun.
  Lett.}, vol.~16, no.~10, pp. 1668--1671, Oct. 2012.

\bibitem{Wang-PCC-polar}
T.~Wang, D.~Qu, and T.~Jiang, ``Parity-check-concatenated polar codes,''
  \emph{IEEE Commun. Lett.}, vol.~20, no.~12, pp. 2342--2345, Dec. 2016.

\bibitem{Trifonov-polar-dynamic-frozen}
P.~Trifonov and V.~Miloslavskaya, ``Polar codes with dynamic frozen symbols and
  their decoding by directed search,'' in \emph{Proc. IEEE Info. Theory
  Workshop}, 2013.

\bibitem{Trifonov-polar-subcode}
------, ``Polar subcodes,'' \emph{IEEE J. Sel. Areas Commun.}, vol.~34, no.~2,
  pp. 254--266, Feb. 2016.

\bibitem{Zhang-PC-polar-Huawei}
H.~Zhang, R.~Li, J.~Wang, S.~Dai, G.~Zhang, Y.~Chen, H.~Luo, and J.~Wang,
  ``Parity-check polar coding for {5G} and beyond,'' in \emph{Proc. IEEE Int.
  Conf. Commun.}, 2018.

\bibitem{arikan-pac}
E.~Ar{\i}kan, ``From sequential decoding to channel polarization and back
  again,'' \emph{arXiv preprint arXiv:1908.09594}, 2019, [Online]. Available:
  https://arxiv.org/abs/1908.09594.

\bibitem{moradi-pac}
M.~Moradi, A.~Mozammel, K.~Qin, and E.~Arikan, ``Performance and complexity of
  sequential decoding of {PAC} codes,'' \emph{arXiv preprint arXiv:2012.04990},
  2020, [Online]. Available: https://arxiv.org/abs/2012.04990.

\bibitem{vardy-pac}
H.~Yao, A.~Fazeli, and A.~Vardy, ``List decoding of {Ar{\i}kan’s PAC}
  codes,'' \emph{Entropy}, vol.~23, no.~7, p. 841, 2021.

\bibitem{Moradi-PAC-Fano-metric}
M.~Moradi, ``On sequential decoding metric function of polarization-adjusted
  convolutional {(PAC)} codes,'' \emph{IEEE Trans. Commun.}, vol.~69, no.~12,
  pp. 7913--7922, Dec. 2021.

\bibitem{Rowshan-PAC-Fano-metric}
M.~Rowshan, A.~Burg, and E.~Viterbo, ``Polarization-adjusted convolutional
  {(PAC)} codes: Sequential decoding vs list decoding,'' \emph{IEEE Trans. Veh.
  Technol.}, vol.~70, no.~2, pp. 1434--1447, Feb. 2021.

\bibitem{Liu-PAC-construction}
W.~Liu, L.~Chen, and X.~Liu, ``A weighted sum based construction of {PAC}
  codes,'' \emph{IEEE Commun. Lett.}, vol.~27, no.~1, pp. 28--31, Jan. 2023.

\bibitem{Rowshan-PAC-construction}
M.~Rowshan, S.~Hoang~Dau, and E.~Viterbo, ``Improving the error coefficient of
  polar codes,'' in \emph{Proc. IEEE Info. Theory Workshop}, 2022, pp.
  249--254.

\bibitem{Moradi-pac-monte-carlo}
M.~Moradi and A.~Mozammel, ``A {Monte-Carlo} based construction of
  polarization-adjusted convolutional {(PAC)} codes,'' \emph{arXiv preprint
  arXiv:2106.08118}, 2021, [Online]. Available:
  https://arxiv.org/abs/2106.08118.

\bibitem{Miloslavskaya-recursive-design}
V.~Miloslavskaya, B.~Vucetic, Y.~Li, G.~Park, and O.-S. Park, ``Recursive
  design of precoded polar codes for {SCL} decoding,'' \emph{IEEE Trans.
  Commun.}, vol.~69, no.~12, pp. 7945--7959, Dec. 2021.

\bibitem{Chiu-PAC-consgruction}
M.-C. Chiu and Y.-S. Su, ``Design of polar codes and {PAC} codes for {SCL}
  decoding,'' \emph{IEEE Trans. Commun.}, vol.~71, no.~5, pp. 2587--2601, May
  2023.

\bibitem{li-pretransformed}
B.~Li, H.~Zhang, and J.~Gu, ``On pre-transformed polar codes,'' \emph{arXiv
  preprint arXiv:1912.06359}, 2019, [Online]. Available:
  https://arxiv.org/abs/1912.06359.

\bibitem{li-pretransformed-average}
Y.~Li, H.~Zhang, R.~Li, J.~Wang, G.~Yan, and Z.~Ma, ``On the weight spectrum of
  pre-transformed polar codes,'' in \emph{Proc. IEEE Int. Symp. Inf. Theory},
  2021, pp. 1224--1229.

\bibitem{Zhou-segmented-CApolar-SCL}
H.~Zhou, C.~Zhang, W.~Song, S.~Xu, and X.~You, ``Segmented {CRC}-aided {SC}
  list polar decoding,'' in \emph{Proc. IEEE Veh. Technol. Conf.}, 2016.

\bibitem{Liang-segmented-BCH-CApolar-SCL}
X.~Liang, H.~Zhou, Z.~Wang, X.~You, and C.~Zhang, ``Segmented successive
  cancellation list polar decoding with joint {BCH-CRC} codes,'' in \emph{Proc.
  Asilomar Conf. Signals, Syst. Comput.}, 2017, pp. 1509--1513.

\bibitem{Gelincik-polar-row-merging}
S.~Gelincik, P.~Mary, J.-Y. Baudais, and A.~Savard, ``Achieving {PAC} code
  performance with {SCL} decoding without extra computational complexity,'' in
  \emph{Proc. IEEE Int. Conf. Commun.}, 2022, pp. 104--109.

\bibitem{costello-road2capacity}
D.~J. Costello and G.~D. Forney, ``Channel coding: The road to channel
  capacity,'' \emph{Proceedings of the IEEE}, vol.~95, no.~6, pp. 1150--1177,
  Jun. 2007.

\bibitem{Tal-polar-construction}
I.~Tal and A.~Vardy, ``How to construct polar codes,'' \emph{IEEE Trans. Inf.
  Theory}, vol.~59, no.~10, pp. 6562--6582, Oct. 2013.

\bibitem{Trifnov-polar-construction}
P.~Trifonov, ``Efficient design and decoding of polar codes,'' \emph{IEEE
  Trans. Commun.}, vol.~60, no.~11, pp. 3221--3227, Nov. 2012.

\bibitem{3gpp-nr-coding}
3GPP, ``{NR}; multiplexing and channel coding,'' \emph{Tech. Rep. TS 38.212,
  Rel. 16}, Jul. 2020.

\bibitem{BOSS-URLLC}
D.~Han, J.~Park, Y.~Lee, H.~V. Poor, and N.~Lee, ``Block orthogonal sparse
  superposition codes for ultra-reliable low-latency communications,''
  \emph{arXiv preprint arXiv:2208.06835}, 2022, [Online]. Available:
  https://arxiv.org/abs/2208.06835.

\end{thebibliography}

\end{document}